\documentclass[twocolumn,showpacs,preprintnumbers,amsmath,amssymb]{revtex4}
\usepackage{graphicx}% Include figure files
\usepackage{dcolumn}% Align table columns on decimal point
\usepackage{bm}% bold math
\usepackage{mathrsfs}
\begin{document}

%\preprint{APS/123-QED}

\title{Electroweak interactions in a relativistic Fermi gas}

\author{K.~Vantournhout}
 \email{Klaas.Vantournhout@UGent.be}
\author{N.~Jachowicz}
\author{J.~Ryckebusch}
\affiliation{
Department of Subatomic and Radiation Physics, Ghent University,\\
Proeftuinstraat 86,\\
B-9000 Gent, Belgium
}
\date{\today}

\begin{abstract}
    
We present a relativistic model for computing the neutrino mean free
path in neutron matter. Thereby, neutron matter is described as a
non-interacting Fermi gas in beta-equilibrium.  We present results for
the neutrino mean free path for temperatures from 0 up to 50 MeV and a
broad range of neutrino energies. We show that relativistic effects
cause a considerable enhancement of neutrino-scattering cross-sections
in neutron matter. The influence of the $Q^2$-dependence in the
electroweak form factors and the inclusion of a weak magnetic term in
the hadron current is discussed. The weak-magnetic term in the hadron
current is at the origin of some selective spin dependence for the
nucleons which are subject to neutrino interactions.
\end{abstract}

\maketitle

\section{Introduction}\label{sec1}
Neutrino interactions play a leading role in a variety of
astrophysical phenomena. For example, neutrinos witnessed nuclear
processes during the Big Bang, they became famous as messengers from
our sun, and are recognized as the representatives from
distant stars. Reaching us from the center of an exploding star,
neutrinos are the first heralds announcing the end of the life of a
massive star \cite{Buras:2003sn}.  Their impact on the dynamics of a
type-II supernova is crucial for the fate of the event.  In the
gravitational collapse of a massive star at the end of its life, an
enormous amount of neutrinos is produced.  During the last stages of
the infall, the neutrinos are trapped inside the star's core.  When
the shock wave passes, densities in the star's center fall and the
neutrinos become free to rush out, carrying away information about the
structure of the proto neutron-star and about the supernova explosion
mechanism \cite{Buras:2003sn}.  For this, the basic mechanisms were
established by the neutrinos observed from supernova 1987A.
Theoretical simulations of core-collapse events \cite{Buras:2003sn},
however, still face some major challenges. The shock wave tends to loose
too much energy when traversing the iron core and no successful
explosions can be reproduced. It has been suggested that the problems
which the numerical simulations are facing, can be attributed to an
incomplete description of the weak interaction processes involved
\cite{Buras:2003sn}.  An excellent knowledge about the cross sections for
neutrino interactions with nuclei and nuclear matter over a wide range
of densities and temperatures is of the utmost importance, hence
constituting one of the major motivations for this work.

In recent years, the interactions of neutrinos with matter at
supra-nuclear densities attracted a lot of attention.  In most
studies, approximations were made with respect to the description of
the hadron gas (degrees of degeneracy, relativity, correlations) and
the weak interaction (hadron current, relativistic effects). Tubbs and
Schramm \cite{Tubbs:1975jx} pioneered calculations of the response of
a hadron gas to electroweak fields.  They investigated the extreme
degenerate (or, quantummechanical) and completely non-degenerate (or,
classical) limits of these processes for a non-interacting Fermi
gas. Schinder \cite{Schinder:1990yr} extended this work to all
temperature and density conditions, thereby adopting an analytical
technique based on the use of Fermi integrals.  Several attempts have
been made to implement the effect of correlations in the description
of the hadron gas.  Burrows and Lattimer \cite{Burrows:1986me} and
Bruenn \cite{Bruenn:1985en} included interactions in the Fermi gas
using effective scaling factors.  Ref.~\cite{Reddy:1997yr} extended
these efforts to a microscopic description of interacting matter of
arbitrary composition and degeneracy, in both non-relativistic and
relativistic frameworks.  Burrows and Sawyer \cite{Burrows:1998cg}
used correlation functions and Horowitz and Wehrberger
\cite{Horowitz:1990it} adopted a relativistic random phase
approximation (RPA) approach.  For all of the above-mentioned
investigations, the weak interaction was modeled as a four-point
Fermi-interaction with a four-momentum $Q^2$ independent hadron current thereby
omitting weak magnetism. The weak magnetic contribution was shown to
be non-negligible in
Ref.~\cite{Vogel:1983hi}. Ref.~\cite{Horowitz:2004yf} included the
weak magnetic contributions within a non-relativistic RPA approach.
Currently, new numerical techniques to investigate dense neutron matter are explored.
Examples include studies based on molecular dynamics techniques
\cite{Watanabe:2003xu,Horowitz:2004yf}, describing the properties of
the 'pasta' phases of nucleon matter and the correlated Tamm-Dancoff approximation (CDTA) \cite{Pandharipande:2006}.

In this paper we study lepton interactions with a non-interacting
relativistic hadron gas, and pay special attention to relativistic
mechanisms. In our approach nucleon matter is described in terms of a
relativistic Fermi gas (RFG).  This approach allows us to investigate the effect of
relativistic kinematics on the phase-space in the
interactions, and its influence on the cross sections. In contrast to
many previous studies, we retain the electroweak current in its full
complexity and introduce $Q^2$-dependent electroweak form factors.

The outline of the paper is as follows.  Section II sketches the RFG
framework which we adopt. Section~III presents the main ingredients of
cross-section calculations for weak interactions between a neutrino
and a nucleon gas.  After testing the reliability of the relativistic
formalism and the adopted numerical techniques in sections
\ref{subsecA} and \ref{limitsec}, relativistic cross-section results
are presented and discussed in sections \ref{sec4:1} and \ref{sec4:2}.

\section{Two-body scattering in a relativistic Fermi gas}
We consider processes in which a particle $a$ is impinging on a
particle $b$, embedded in a gas.  The differential cross section for
the collision of the two particles resulting in the creation of $N$
reaction products $a+b\rightarrow\sum_{f=1}^N x_f$ is given by
\begin{multline}\label{eq211}
d^{3N}\sigma = \frac{(2\pi)^{10}}{v_{rel}}\prod_f
d^3\vec{p}_f\delta^4(p_a+p_b-\sum_fp_f) \\
\times\left|M(a+b\rightarrow\{f\})\right|^2,
\end{multline}
where $p_a$ and $p_b$ are the four-momenta of the colliding particles
and the $N$ outgoing particles $f$ have four-momentum $p_f$. In a
relativistic description, energy and momentum conservation are imposed
through the four-dimensional Dirac delta function
$\delta^4(p_a+p_b-\sum_fp_f)$.  The dynamics of the interaction is
described by the transition matrix element $M(a+b\rightarrow\{f\})$.
The relative velocity $v_{rel}$ represents the flux of incident
particles.  The general Lorentz-covariant formulation of this quantity
is given by
\begin{equation}\label{vel}
v_{rel} = \frac{\sqrt{(p_ap_b)^2-p_a^2p_b^2}}{p_a^0p_b^0} \; .
\end{equation}
For head-on collisions, Eq.~\eqref{vel} reduces to the familiar expression
\begin{equation}
v_{rel}=\left|\frac{\vec{p}_a}{p_a^0} - \frac{\vec{p}_b}{p_b^0}\right|.
\end{equation}

The gas of particles $b$ on which the particles $a$ are impinging is
modeled in terms of the probability distribution
$F^b_{\alpha_b}(\vec{r}_b,\vec{p}_b)$, describing the probability that
a target particle $b$ with quantum numbers $\alpha_b$ occupies a
phase-space volume $d^3\vec{r}_bd^3\vec{p}_b$.  For a homogeneous and non-interacting
gas, the probability distributions are independent of $\vec{r}_b$:
$F^b_{\alpha_b}(\vec{r}_b,\vec{p}_b) = F^b_{\alpha_b}(\vec{p}_b)$. For
fermions, the cross section of Eq.~\eqref{eq211} can be cast in the form
\begin{widetext}
\begin{equation}\label{eq212}
\mathcal{N} d^{3N}\bar\sigma = \sum_{\substack{\alpha_a,\alpha_b,\\
\{\alpha_f\}}}\int
F^b_{\alpha_b}(\vec{p}_b)d^3\vec{p}_b\:\frac{(2\pi)^{10}}{v_{rel}}\prod_f
F'^{f}_{\alpha_f}(\vec{p}_f)
d^3\vec{p}_f\delta^4(p_a+p_b-\sum_fp_f)\left|M(a+b\rightarrow\{f\})\right|^2.
\end{equation}
\end{widetext}
Here, $d^{3N}\bar\sigma$ represents the differential cross section per
particle $b$, $F'^{f}_{\alpha_f}(\vec{p}_f)$ is the probability that
the final state determined by $(p_f,\alpha_f)$ is unoccupied
\begin{equation}
F'^{x_f}_{\alpha_f}(\vec{p}_f) = 1-F^{x_f}_{\alpha_f}(\vec{p}_f) \; .
\end{equation}
Further, $\mathcal{N}$ is proportional to the number density of the gas 
\[\mathcal{N}= \sum_{\alpha_b}\int
F^b_{\alpha_b}(\vec{p}_b)d^3\vec{p}_b \; ,\] where $\sum_{\alpha_b}$
represents the sum over all relevant quantum numbers.  For an
interaction of the type $ a + b \rightarrow x + y$, the differential
cross section per target particle $b$ can be cast in the form
\begin{widetext}
\begin{align}\label{eq213}
\mathcal{N}\ \frac{d^3\bar\sigma}{d|\vec{p}_x|d^2\Omega_x}
&=\sum_{\substack{\alpha_a,\alpha_b,\\ \alpha_x,\alpha_y}}\int
F^b_{\alpha_b}(\vec{p}_b)d^3\vec{p}_b\:\frac{(2\pi)^{10}}{v_{rel}}\int
F'^y_{\alpha_y}(\vec{p}_y) d^3\vec{p}_y \notag \\ &\qquad\qquad\times
\delta^4(p_a+p_b-p_x-p_y)\left|M(a + b\rightarrow x + y)\right|^2
F'^x_{\alpha_x}(\vec{p}_x) |\vec{p}_x|^2,
\end{align}
\end{widetext}
where it was assumed that both $x$ and $y$ are embedded in a Fermi gas.

For the reaction $a + b \rightarrow x + y$ only a limited region in
phase-space is accessible for the outgoing particles $x$ and $y$.  At
the same time, an outgoing particle $x$ with a four-momentum $p_x$,
can only be created from a well-defined part of the initial phase
space.  The kinematic restrictions governing the interaction are
dictated by four-momentum conservation.  Defining the four-momentum
difference $w = p_a - p_x$ and $W^2 = m_y^2-m_b^2-w^2$, it becomes
straightforward to simplify the integral in Eq.~\eqref{eq213}.  When the labels $a$ and $x$ are  assigned to the incoming and outgoing lepton, the four-momentum difference
$w$ represents the four-momentum exchange $q$.  For $|\vec{w}|\ne 0$, 
Eq.~\eqref{eq213} can be cast in the form
\begin{widetext}
\begin{equation}
\begin{split}
\mathcal{N} \frac{d^3\bar\sigma}{d|\vec{p}_x|d^2\Omega_x} &=
\frac{1}{|\vec{w}|} \sum_{\substack{\alpha_a,\alpha_b,\\
\alpha_x,\alpha_y}} \int\limits_{p_b^{MIN}}^{p_b^{MAX}}
|\vec{p}_b|d|\vec{p}_b|\:\int\limits_0^{2\pi}d\phi_b
F^b_{\alpha_b}(\vec{p}_b) F'^y_{\alpha_y}(\vec{w}+\vec{p}_b)
(w^0+p_b^0) \Theta(w^0+p_b^0-m_y) \notag \\ &\qquad\qquad\times
\frac{(2\pi)^{10}}{v_{rel}}\left|M(a + b \rightarrow x + y) \right|^2
F'^x_{\alpha_x}(\vec{p}_x) |\vec{p}_x|^2,
\end{split}
\end{equation}
\end{widetext}
in which the  condition 
\begin{equation}
\cos\theta_{\vec{w},\vec{p}_b}=\frac{-W^2+2w^0p_b^0}{2|\vec{w}||\vec{p}_b|} = \frac{-W^2+2w_0\sqrt{m_b^2+|\vec{p}_b|^2}}{2|\vec{w}||\vec{p}_b|},
\end{equation}
must be obeyed.  The integration limits for the initial momentum
$|\vec{p}_b|$ are the solutions of $\cos\theta_{\vec{w},\vec{p}_b} =
\pm 1$ and are listed in Table \ref{table1}.
%
% Start of the Table
%
\begin{table*}
\caption{\label{table1} Overview of the integration limits for $|\vec{p}_b|$.}
\begin{ruledtabular}
\begin{tabular}{l|cc}
& $p_b^{MIN}$ & $p_b^{MAX}$ \\
\hline
$w^2<0$ & $\frac{1}{2}\left|\frac{W^2|\vec{w}|}{w^2}+ w_0\sqrt{\frac{W^4}{w^4}-\frac{4m_b^2}{w^2}}\right|$ & $\infty$ \\
$w^2<0,W^4 = 4w^2m_b^2$ & \multicolumn{2}{l}{no solution} \\
$w^2<0,W^2 = 2w_0m_b$ & $0$ & $\infty$ \\
\hline
$w^2=0$ & $\left|\frac{|\vec{w}|m_b^2}{W^2}-\frac{W^2}{4|\vec{w}|}\right|$ & $\infty$ \\
\hline
$w^2>0,W^4>4w^2m_b^2$ & $\frac{1}{2}\left|\left|\frac{W^2|\vec{w}|}{w^2}\right|- |w_0|\sqrt{\frac{W^4}{w^4}-\frac{4m_b^2}{w^2}}\right|$ & $\frac{1}{2}\left|\left|\frac{W^2|\vec{w}|}{w^2}\right|+ |w_0|\sqrt{\frac{W^4}{w^4}-\frac{4m_b^2}{w^2}}\right|$ \\
$w^2>0,W^4 \le 4w^2m_b^2$ & \multicolumn{2}{l}{no solution} \\
$w^2>0,W^4 = 4w_0^2m_b^2$ & $0$ & $\left|\frac{W^2|\vec{w}|}{w^2}\right|$ \\
\end{tabular}
\end{ruledtabular}
\end{table*}\\
%
% End of the table
%

For the peculiar situation that $|\vec{w}| = 0$, Eq.~\eqref{eq213} becomes
\begin{widetext}
\begin{align}
\mathcal{N} \frac{d^3\bar\sigma}{d|\vec{p}_x|d^2\Omega_x} &=
\left|\frac{-W^2}{w_0\sqrt{W^4-4w_0^2m_b^2}}\right|
 \sum_{\substack{\alpha_a,\alpha_b,\\
\alpha_x,\alpha_y}}\iint\limits_{4\pi} d^2\Omega_b F^b_{\alpha_b}(\vec{p}_b)
F'^{y}_{\alpha_y}(\vec{p}_b) (w^0+p_b^0) \Theta(w^0+p_b^0-m_y)\notag\\
&\qquad\qquad\times \frac{(2\pi)^{10}}{v_{rel}}\left|M(a + b \rightarrow
x+y ) \right|^2 F'^x_{\alpha_x}(\vec{p}_x) |\vec{p}_x|^2 |\vec{p}_b|^2,
\end{align}
\end{widetext}
with
\begin{equation}
|\vec{p}_b| = \sqrt{\frac{W^4}{4w_0^2}-m_b^2} \; .
\end{equation}

Nucleon matter at supra-nuclear densities can be described as a
partially degenerate Fermi gas. In the remainder of this paper, the
probability distributions $F^i_{\alpha_i}$ are replaced by Fermi
distributions
\begin{equation}
F^i_{\alpha_i}(\vec{p}) = \frac{1}{\exp\left(\cfrac{E_i(\vec{p})-\mu_i}{kT}\right)+1}.
\end{equation}
The introduction of this probability distribution violates the Lorentz
invariance of Eq.~\eqref{eq213}.

We deal with a mixture of a proton, neutron, electron and neutrino
 gas.  In an equilibrium situation, the second law of thermodynamics
 imposes relations between their respective chemical potentials. For
 reactions of the type
\begin{equation}
\nu_e + n  \leftrightarrow e^- + p,
\end{equation}
the sum of the chemical potentials of the initial and final states
should be equal.  Considering that the total system is neutral and
denoting the baryon density by $n_B$, beta-equilibrium requires 
that the following conditions are obeyed~:
\begin{subequations}
\begin{gather}
\mu_{\nu_e} + \mu_n = \mu_{e^-} + \mu_p,\\
n_p = n_{e^-},\\
n_p + n_n = n_B.
\end{gather}
A fourth condition is related to  neutrino trapping. In the absence of trapping one has 
\begin{gather}
\begin{align}
\mu_{\nu_e} &= 0 \; .
\end{align}
\end{gather}
When trapping occurs one defines
\begin{align}
\frac{n_{e^-}+n_{\nu_e}}{n_B} &= Y_\ell.
\end{align}
\end{subequations}
Common values for $Y_\ell$ are in the range between 0.3 and 0.4
\cite{Watanabe:2005qt}. For fermions, the number density $n_i$ is
determined from
\begin{equation}
n_i = \frac{1}{2\pi^2}\int\limits_0^\infty
\frac{p^2}{\exp\left(\frac{\sqrt{p^2+M_i^2}-\mu_i}{kT}\right)+1}dp \; .
\end{equation}

\section{Electroweak interaction matrix elements}\label{sec2:subsec3}
In first-order perturbation theory, electroweak interactions between
leptons and hadronic matter are described as the four-product of a
lepton $\jmath_{\mu}$ and hadron current $J_{\nu}$, a tensor
$B^{\mu\nu}$ representing the exchanged boson mediator, and a strength
factor $\chi$. We define $p$ ($P$) and $k$ ($K$) as the four
momentum of the initial (final) lepton and hadron, $q$ as the four-momentum exchange, $Q^2$ equals $-q_{\mu}q^{\mu}$.  The chiralities of
the initial and final leptons are denoted by $\lambda$ and
$\Lambda$. Further, $s$ and $S$ are the spins of the initial and final
hadrons. The transition matrix element takes on the general form
\begin{equation}
iM \left( p \lambda + k s \rightarrow P \Lambda + K S \right) = \chi \jmath_\mu B^{\mu\nu} J_\nu, 
\end{equation}
and for neutral current (NC) reactions one has
\begin{widetext}
\begin{equation}
 \jmath_\mu B^{\mu\nu} J_\nu  = 
-i\frac{G_F}{\sqrt{2}}M_Z \langle
P;\Lambda|\jmath_\mu^{NC}|p;\lambda\rangle\frac{g^{\mu\nu}-\frac{q^\mu
q^\nu}{M_Z^2}}{M_Z^2+Q^2}\langle
K;S|J_\nu^{NC}|k;s\rangle,
\end{equation}
for charged current (CC) reactions
\begin{equation}
\jmath_\mu B^{\mu\nu} J_\nu  = 
-i\frac{G_F\cos\theta_C}{\sqrt{2}}M_W \langle
P;\Lambda|\jmath_\mu^{CC}|p;\lambda\rangle\frac{g^{\mu\nu}-\frac{q^\mu
q^\nu}{M_W^2}}{M_W^2+Q^2}\langle
K;S|J_\nu^{CC}|k;s\rangle \; .
\end{equation}
\end{widetext}
In the matrix element for charged-current interactions an extra factor
$\cos\theta_C$ appears.  This is due to the fact that the weak
left-handed $d$-quark is an admixture of the strong left-handed $d$
and $s$ quarks.  As a consequence, strangeness-conserving hadronic
charged-current processes are weaker than their lepton counterparts.

The lepton weak neutral current is generally written as
\begin{equation}
\langle P;\Lambda|\jmath_\mu^{NC}|p;\lambda\rangle = 
4\langle P;\Lambda|\bar\psi\gamma_\mu(I_z^W-\sin^2\theta_W
Q)\psi|p;\lambda\rangle \; ,
\end{equation}
where $I_z^W$ represents the weak isospin operator, $Q$ the charge
operator and $\psi$ are the normalized Dirac field operators.

For neutrinos the currents are given by
\[\langle
P;\Lambda|\jmath_\mu^{CC}|p;\lambda\rangle=\langle
P;\Lambda|\bar\psi(1-\gamma_5)\gamma_\mu\psi|p;\lambda\rangle
,\]
where the projection operator $(1-\gamma_5)$ takes into account
that only left-handed neutrinos couple to the weak interaction.  The
final lepton $\ell$ is a massive lepton $\ell^-$ in CC reactions and a
neutrino $\nu_\ell$ for NC scattering processes.

The weak nucleon current is given by \cite{Bernstein:1968}
\begin{multline}
J_\nu = F_1(Q^2)\gamma_\nu + \frac{i}{2M_N}F_2(Q^2)\sigma_{\nu\mu}q^\mu \\
+ G_A(Q^2)\gamma_\nu\gamma_5 +\frac{1}{2M_N}G_P(Q^2)q_\nu\gamma_5,
\end{multline}
where second-class currents were not considered. $F_1(Q^2)$ and
$F_2(Q^2)$ are the Dirac form factors which are related to the
well-known weak-electric and weak-magnetic form-factors through the
relations $G_E = F_1 - \tau F_2$ and $G_M = F_1 + F_2$ where $\tau =
Q^2/4M_N^2$.  The $Q^2$-dependence of the weak Sachs vector form
factors $G_E$ and $G_M$ is given by the dipole parametrization:
\begin{subequations}
\begin{widetext}
\begin{gather}
G(Q^2) = G(0)\left(1-\cfrac{Q^2}{M_V^2}\right)^{-2},\qquad M_V^2 =
710649 \text{ MeV}^2\\ G_E^0(0) =
\left(\frac{1}{2}-\sin^2\theta_W\right)\tau_3 - \sin^2\theta_W,\qquad
G_E^{\pm}(0) = 1\\ G_M^0(0) =
\left(\frac{1}{2}-\sin^2\theta_W\right)(\mu_p-\mu_n)\tau_3-\sin^2\theta_W(\mu_p+\mu_n),\qquad
G_M^{\pm}(0) = (\mu_p+\mu_n)
\end{gather}
where $\sin^2\theta_W=0.2312$, $\mu_p$ and $\mu_n$ are the proton and
neutron magnetic moments and $\tau_3$ is the isospin operator.  The
axial form factor is given by
\begin{gather}
G^{\pm}_A(Q^2) = -g_a\left(1-\cfrac{Q^2}{M_a^2}\right)^{-2},\qquad G^{0}_A(Q^2) = -\frac{g_a}{2}\left(1-\cfrac{Q^2}{M_a^2}\right)^{-2}\\
g_a = 1.262,\qquad M_a^2 = 1065024 \text{ MeV}^2.
\end{gather}
\end{widetext}
\end{subequations}
The pseudoscalar form factor is related to the axial one through the Goldberger-Treiman relation
\begin{equation}
G_P(Q^2) = -\frac{4M^2 G_A(Q^2)}{Q^2+m_\pi^2}.
\end{equation}
As $q_\mu$ can be written in terms of the lepton masses, the
pseudoscalar contribution vanishes for neutral-current processes.

\section{Results}
Before presenting the numerical results of the cross-section
calculations for neutrino interactions with nucleon matter in sections
\ref{sec4:1} and \ref{sec4:2}, we establish the reliability of our
formalism and the adopted numerical techniques. To this end, we study
some well-defined limits for which our predictions can be compared with
results from alternative approaches.

\subsection{Neutrino scattering on a single nucleon}\label{subsecA}
In former studies of the neutrino-nucleon and neutrino-nucleus
response, a variety of expressions for the hadron currents has been
used~:
\begin{eqnarray}
J_\nu^{V-A} &=& (1-\gamma_5)\gamma_\nu,\label{cur1} \\ 
J_\nu^{RN} &=&
(c_v - c_a\gamma_5)\gamma_\nu,\label{cur2} \\ 
J_\nu^{Mag} &=&
F_1(0)\gamma_\nu + \frac{i}{2M_N}F_2(0)\sigma_{\nu\mu}
q^\mu  \nonumber  \\ 
&&\qquad + G_A(0)\gamma_\nu\gamma_5 +
\frac{1}{2M_N}G_P(0)q_\nu\gamma_5,\label{cur3} \\ 
J_\nu^{Q^2} &=&
F_1(Q^2)\gamma_\nu + \frac{i}{2M_N}F_2(Q^2)\sigma_{\nu\mu} q^\mu
\nonumber \\ 
&&\qquad + G_A(Q^2)\gamma_\nu\gamma_5 +
\frac{1}{2M_N}G_P(Q^2)q_\nu\gamma_5.\label{cur4}
\end{eqnarray}

\begin{figure}
\includegraphics[width=0.49\textwidth]{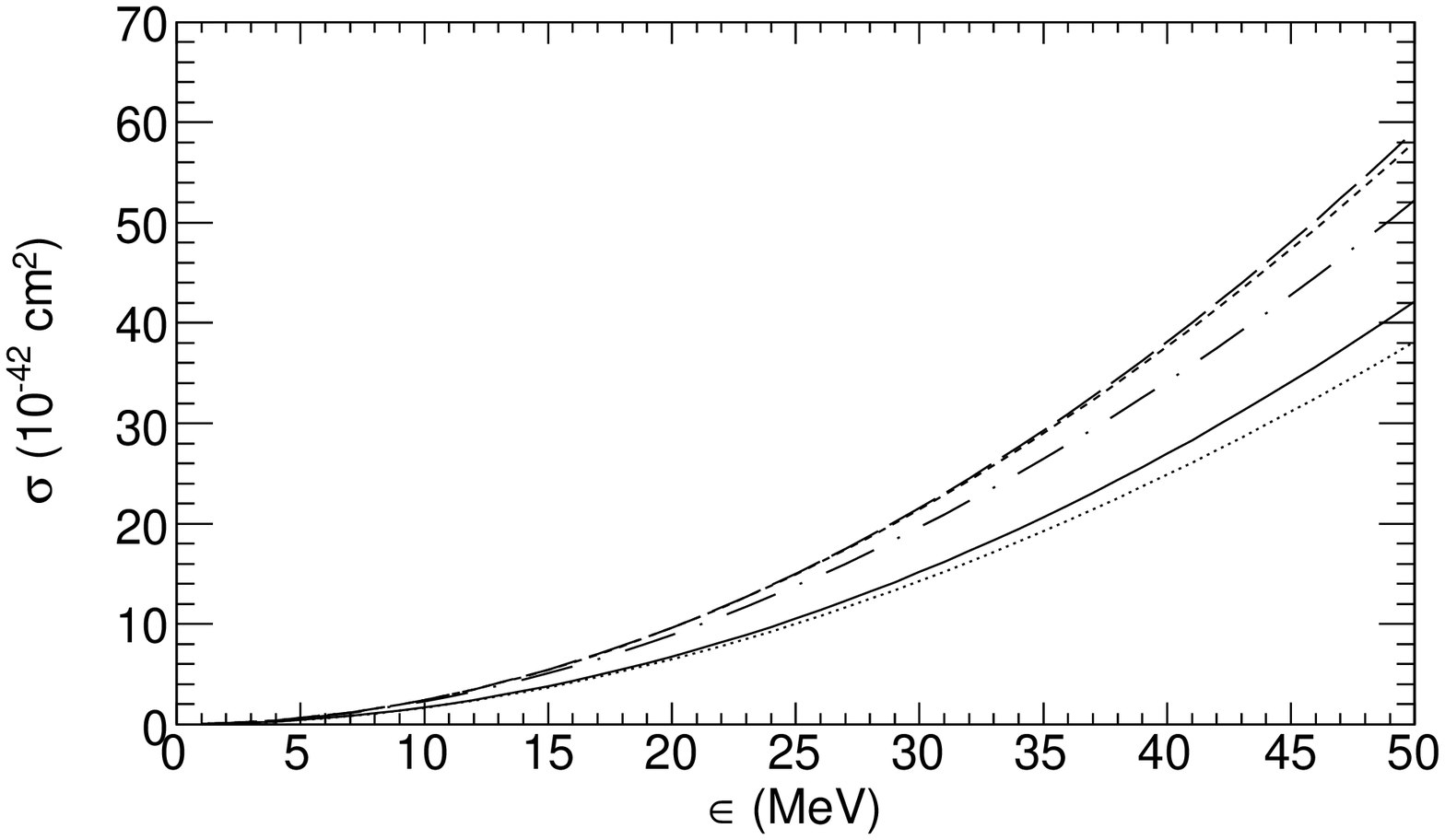}\hspace{\stretch{1}}
\includegraphics[width=0.49\textwidth]{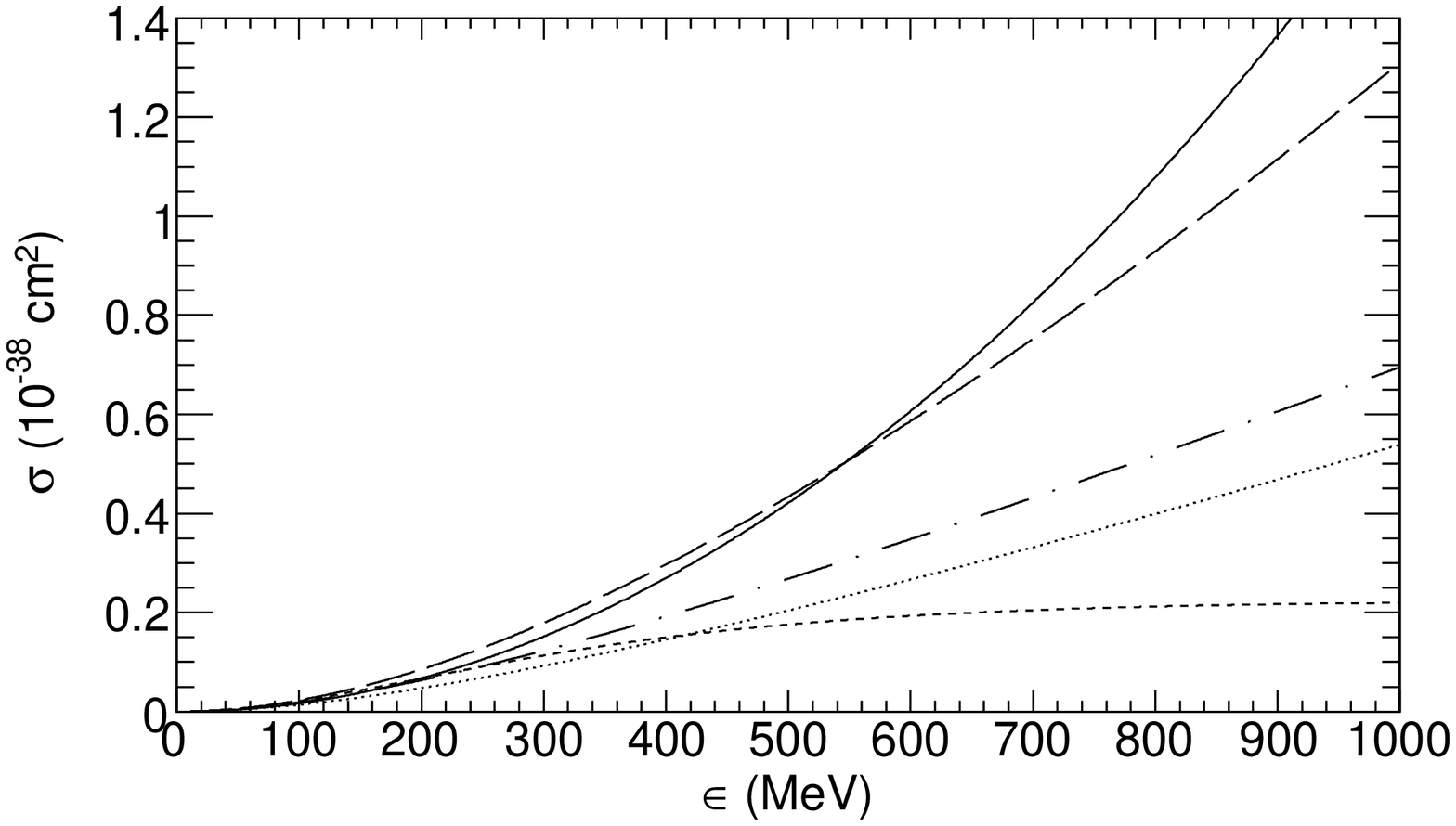}
\caption{The neutrino-neutron cross sections as a function of the
  incoming neutrino energy $\epsilon$ as obtained
with different expressions for the nucleon current. The solid line
  refers to the analytical approximation of the cross section according to Eq.~\eqref{sigmatubbs} ; the dotted line to cross sections obtained using the 
V-A current of Eq.~\eqref{cur1} ;  the renormalization effects of Eq.~\eqref{cur2} are additionally taken into account  in the dash-dotted curve. The long-dashed line corresponds to the cross sections for the current operator of Eq.~\eqref{cur3} where the weak-magnetic contribution was added. The $Q^2$-dependence of the form factors shown in Eq.~\eqref{cur4} is taken into account in the short-dashed curve.}
\label{onenucl}
\end{figure}

The operator $J_\nu^{V-A}$ of Eq.~\eqref{cur1} assumes that hadrons are
point-like particles, with the weak interaction occurring on the
hadron instead of on the quark level.  With this hadron current, an approximation for the neutrino-nucleon cross section was derived in Ref.~\cite{Tubbs:1975jx} :
\begin{equation}
\sigma^{T-Schr}(\epsilon)=\frac{1}{4}\frac{4G_F^2m_e^2\hbar^2}{\pi
  c^2}\left(\frac{\epsilon}{m_e}\right)^2 \;.\label{sigmatubbs}
\end{equation}
This expression is valid when the incoming neutrino-energy is sufficiently small in
comparison with the nucleon mass $\epsilon \ll M_N$.  The validity of this approximation can be appreciated from Fig.~\ref{onenucl}, when comparing the solid line, representing the approximation \eqref{sigmatubbs}, and the dotted line representing the cross section calculated with the current \eqref{cur1}.  From the top panel it emerges that
the analytical expression of Eq.~\eqref{sigmatubbs} is fine at low energies. From
the results in the lower panel it is clearly seen that the
approximation \eqref{sigmatubbs} is not valid any more at higher neutrino energies.
The current
operator $J_\nu^{RN}$ of Eq.~\eqref{cur2} introduces renormalization effects that can be
attributed to the strong interaction and the quark structure of the
hadron.  In the expression $J_\nu^{Mag}$, the weak magnetic and
pseudo-scalar contributions are added. The pseudo-scalar term
vanishes for neutral-current processes.  Due to its dependence on the
momentum transfer, the contribution of the magnetic term is small at
low energies.  The operator $J_\nu^{Q^2}$ exhibits a
$Q^2$-dependence in the form factors and can be interpreted as the
most accurate expression for the hadronic weak current to date.

For each of the above current operators we display the computed
neutrino-neutron cross section in Fig.~\ref{onenucl}. Thereby, we
discriminate between low neutrino energies (up to 50 MeV) and neutrino
energies that go up to the nucleon mass. At small incoming neutrino
energies, our numerical results are in good agreement with the
analytical expression of Eq.~\eqref{sigmatubbs}.  With increasing
neutrino energies, the deviation between the expression
Eq.~\eqref{sigmatubbs} and results obtained with more sophisticated
current operators grows. The magnetic contribution in the current operator
considerably enhances the cross sections at higher energies, while the
introduction of $Q^2$-dependent form factors reduces the
cross section strongly at large incoming neutrino energies.

\subsection{Neutrino interactions with a neutron gas in the ($T=0$ K, $\mu=M_n$ MeV) limit}\label{limitsec}

\begin{figure*}
\includegraphics[width=0.49\textwidth]{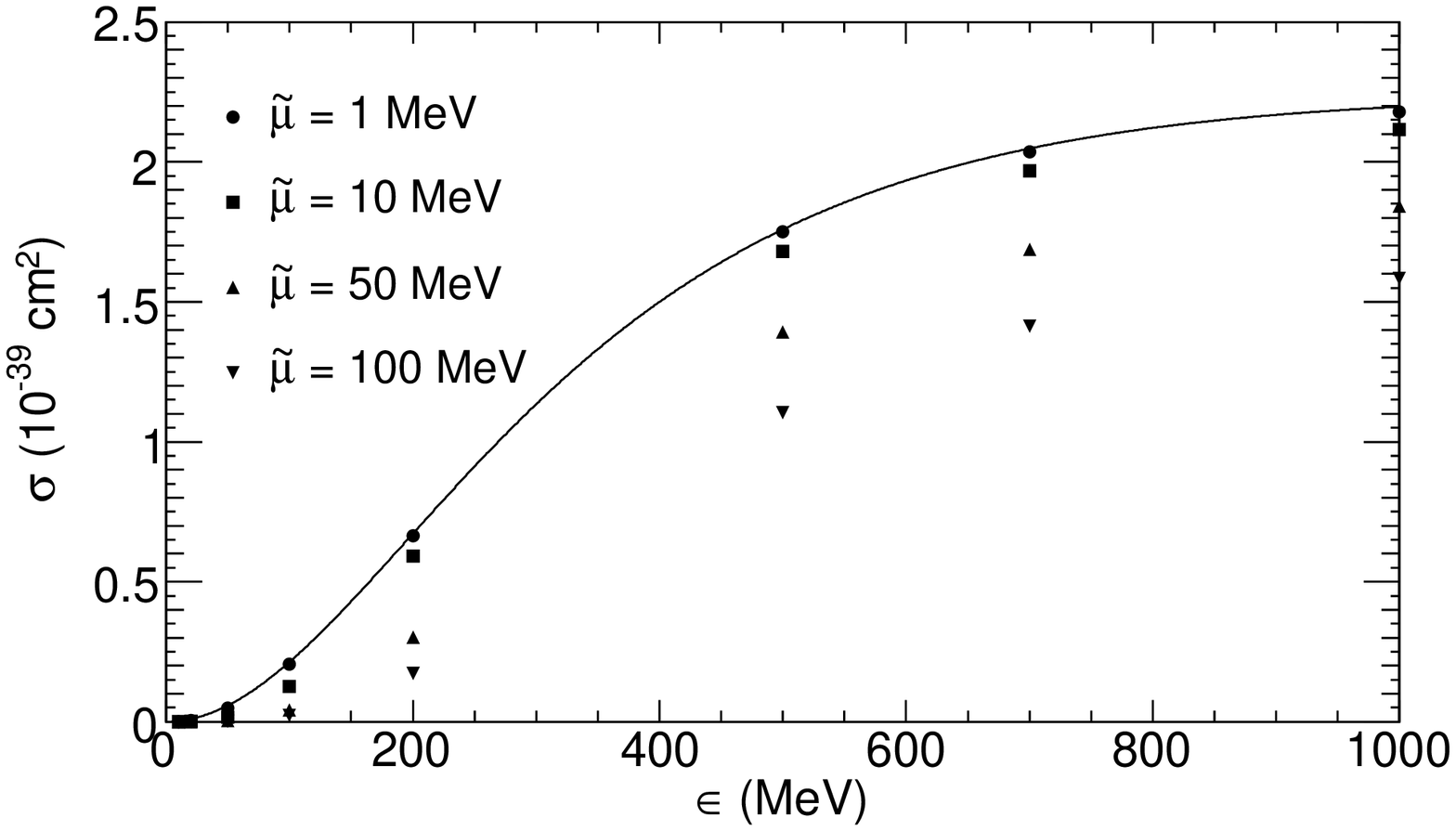}\hspace{\stretch{1}}
\includegraphics[width=0.49\textwidth]{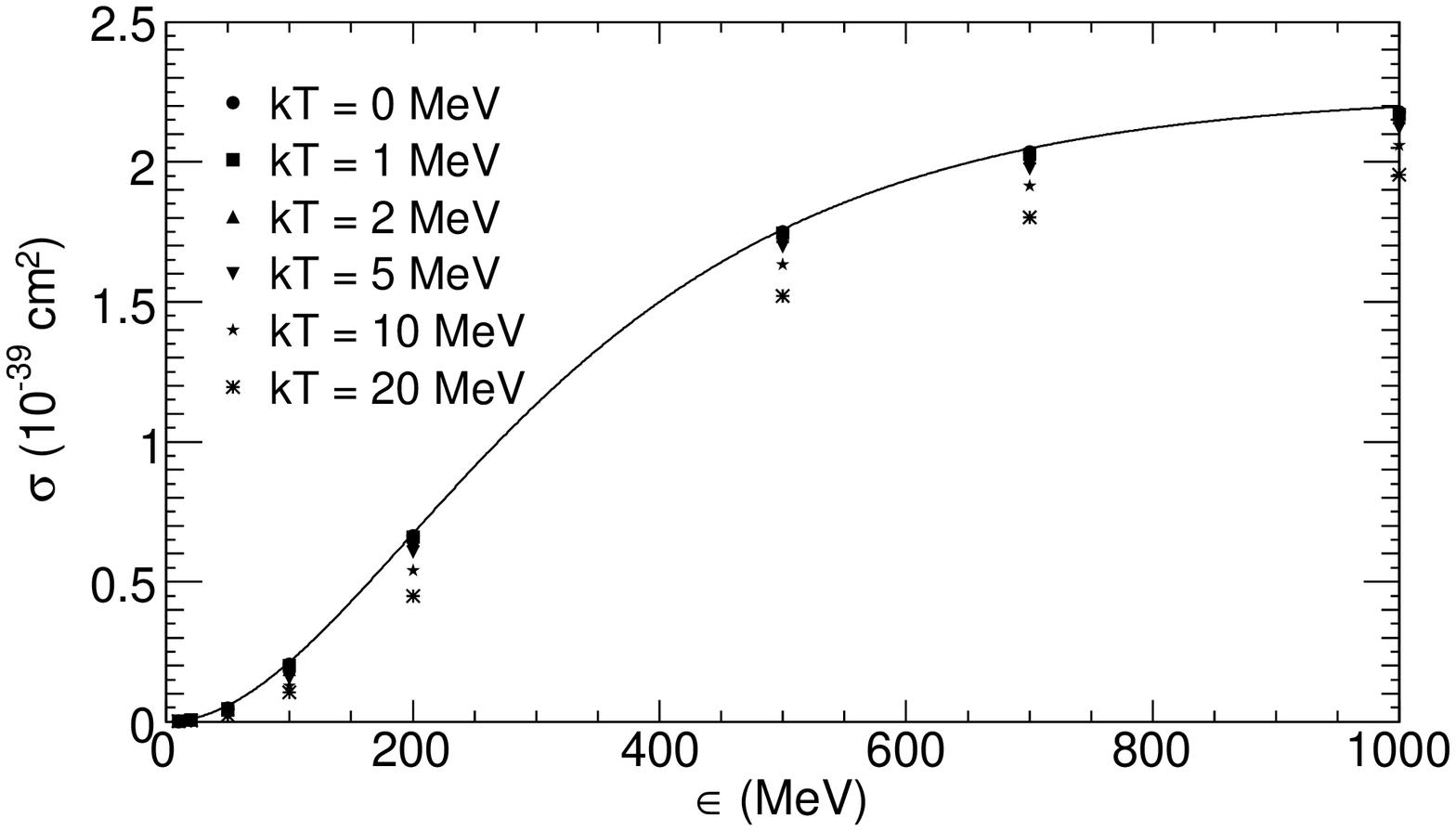}
\caption{Comparison between neutral-current cross sections on a single
nucleon (solid line) and in a Fermi gas (markers
). The left panel displays
results for different values of the chemical potential and $T = 0$ K.
In the right panel the temperature dependence is investigated for a
chemical potential of $\tilde\mu = 1$ MeV.  The chemical potentials are
corrected for the rest mass of the free
neutron.}\label{single_fermi_compare}
\end{figure*}

When the value of the chemical potential $\mu$ approaches the rest mass of a
free nucleon, or $\tilde\mu = \mu - M_n$ approaches zero and the temperature comes close to $T=0$ K,
neutrino-scattering cross sections in the neutron gas are expected to
approach those of neutrino interactions on a single neutron in its
rest frame.  Indeed, at $T=0$ K a decrease of the chemical potential
implies a reduction of the average kinetic energy. At $\mu=M_n$, the
kinetic energy vanishes.  Hence, this limit of the Fermi-gas
calculation provides a stringent test of the formalism and of the
adopted numerical techniques.  The results of this comparison between
Fermi-gas and free-neutron cross sections, are presented in
Fig.~\ref{single_fermi_compare}. When turning off the effect of the
chemical potential ($\tilde\mu\rightarrow 0$), the influence of the other particles on the struck
nucleon diminishes, a process which results in growing cross sections.
The figure demonstrates that in the limit of vanishing chemical
potential the cross sections computed in the hadron gas coincide with those on a single nucleon.
The right panel of Fig.~\ref{single_fermi_compare} shows the
temperature dependence of the RFG prediction at chemical potentials
approaching the rest mass of the hadrons.  The numerical calculations
were performed for $\tilde\mu =  1$ MeV. Again, the results converge to
the single-nucleon cross-sections in the $T=0$ K limit.  This
observation provides convincing evidence for the reliability of the
adopted theoretical framework.

\subsection{Relativistic effects in the phase-space factor}\label{sec4:1}
\begin{figure}
\includegraphics[scale=0.45]{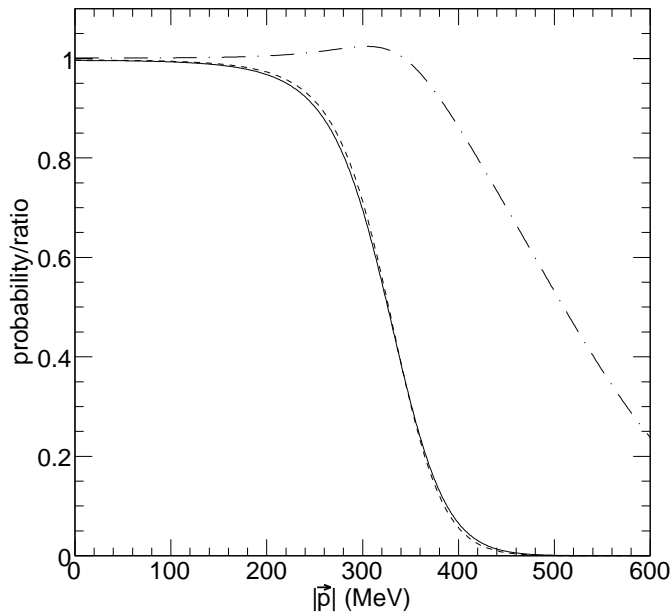}
\caption{Comparison between a relativistic ($F_{R}$, solid line) and a
non-relativistic Fermi distribution ($F_{NR}$, dashed line), for $kT = 10$ MeV and $n =
0.16$ fm$^{-3}$.  The dash-dotted line represents the ratio $F_{NR} /
F_{R} $.}\label{fermicompare}
\end{figure}

\begin{figure*}
\includegraphics[width=0.49\textwidth]{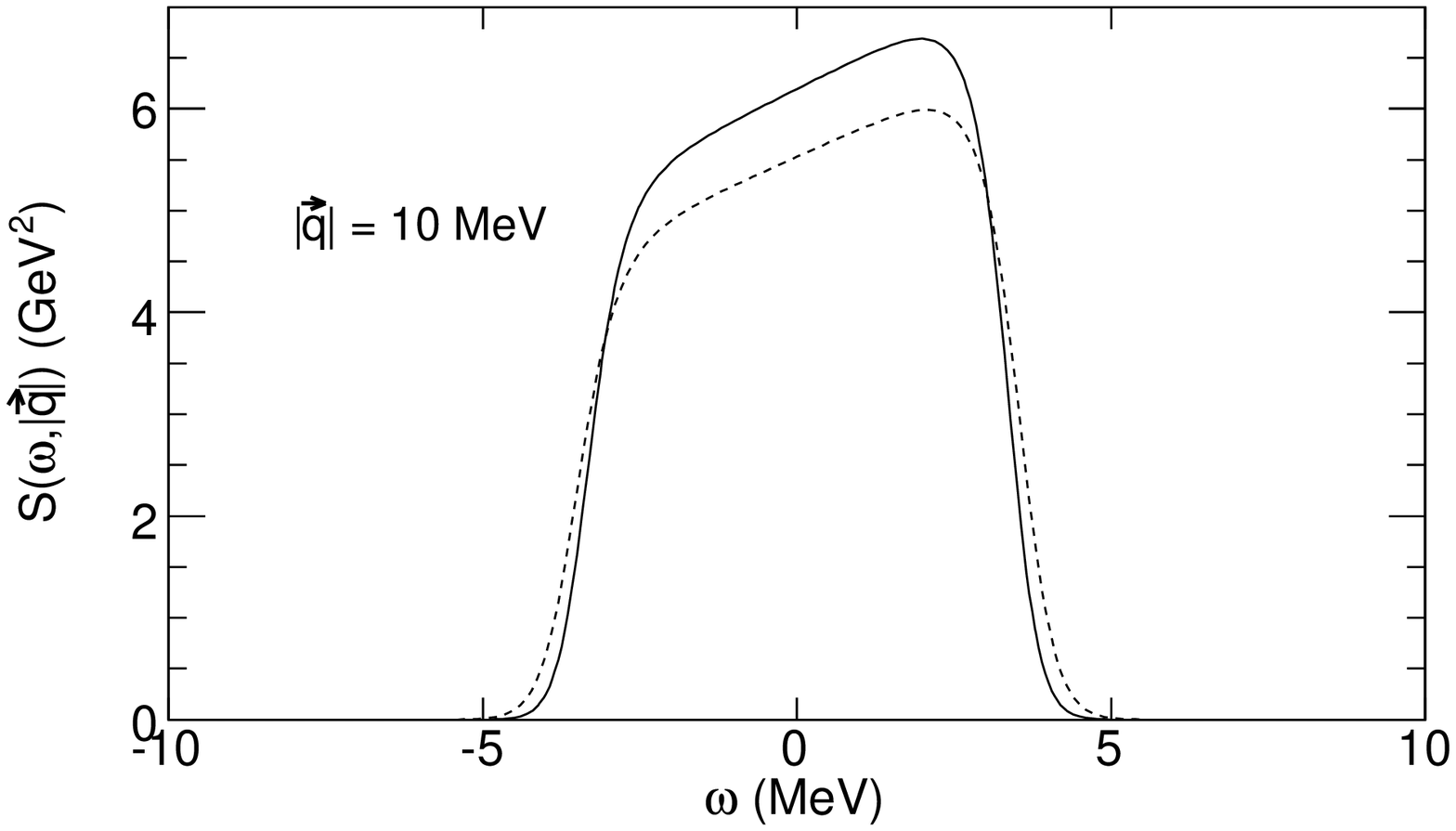}\hspace{\stretch{1}}
\includegraphics[width=0.49\textwidth]{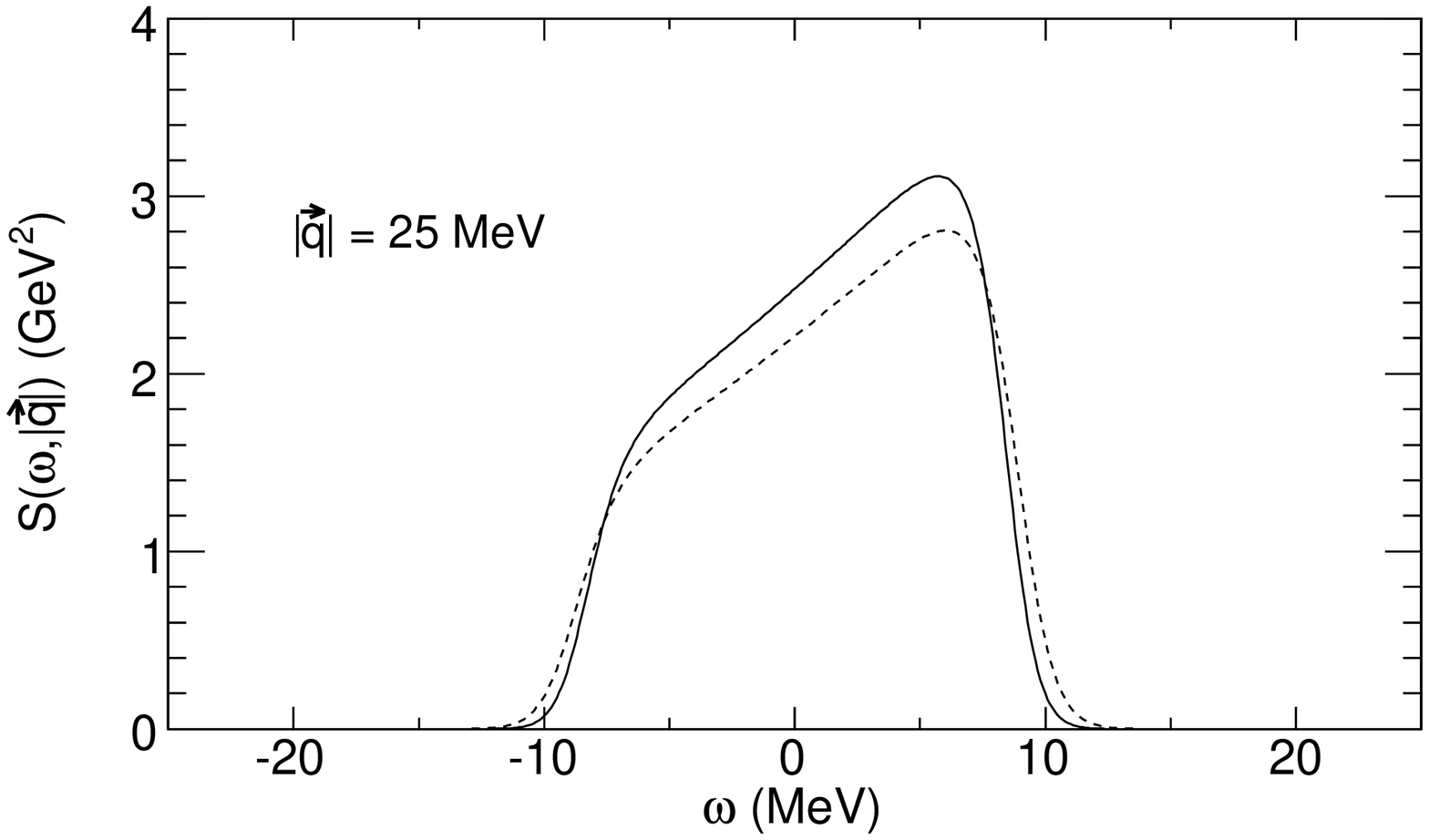}\\
\includegraphics[width=0.49\textwidth]{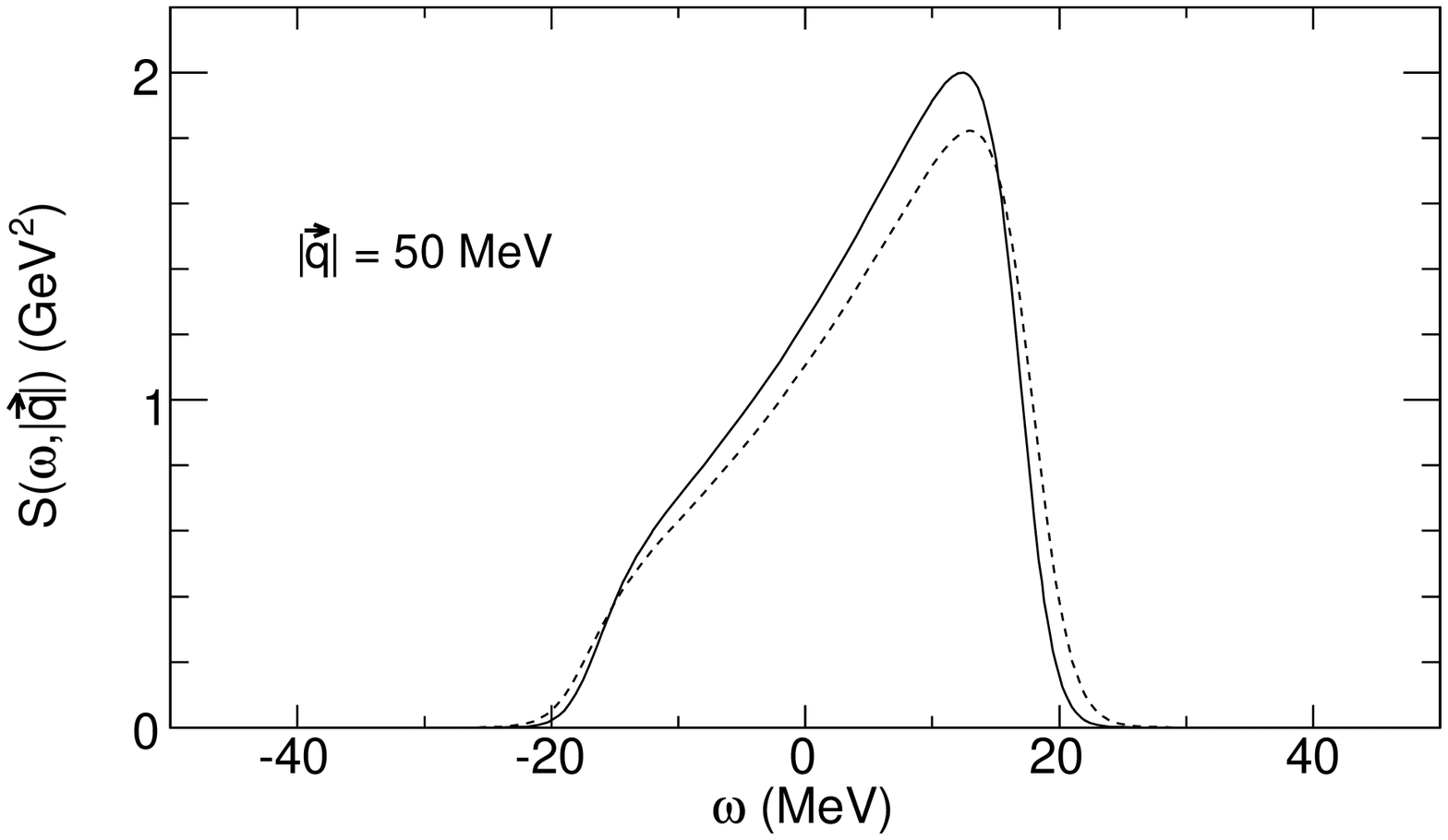}\hspace{\stretch{1}}
\includegraphics[width=0.49\textwidth]{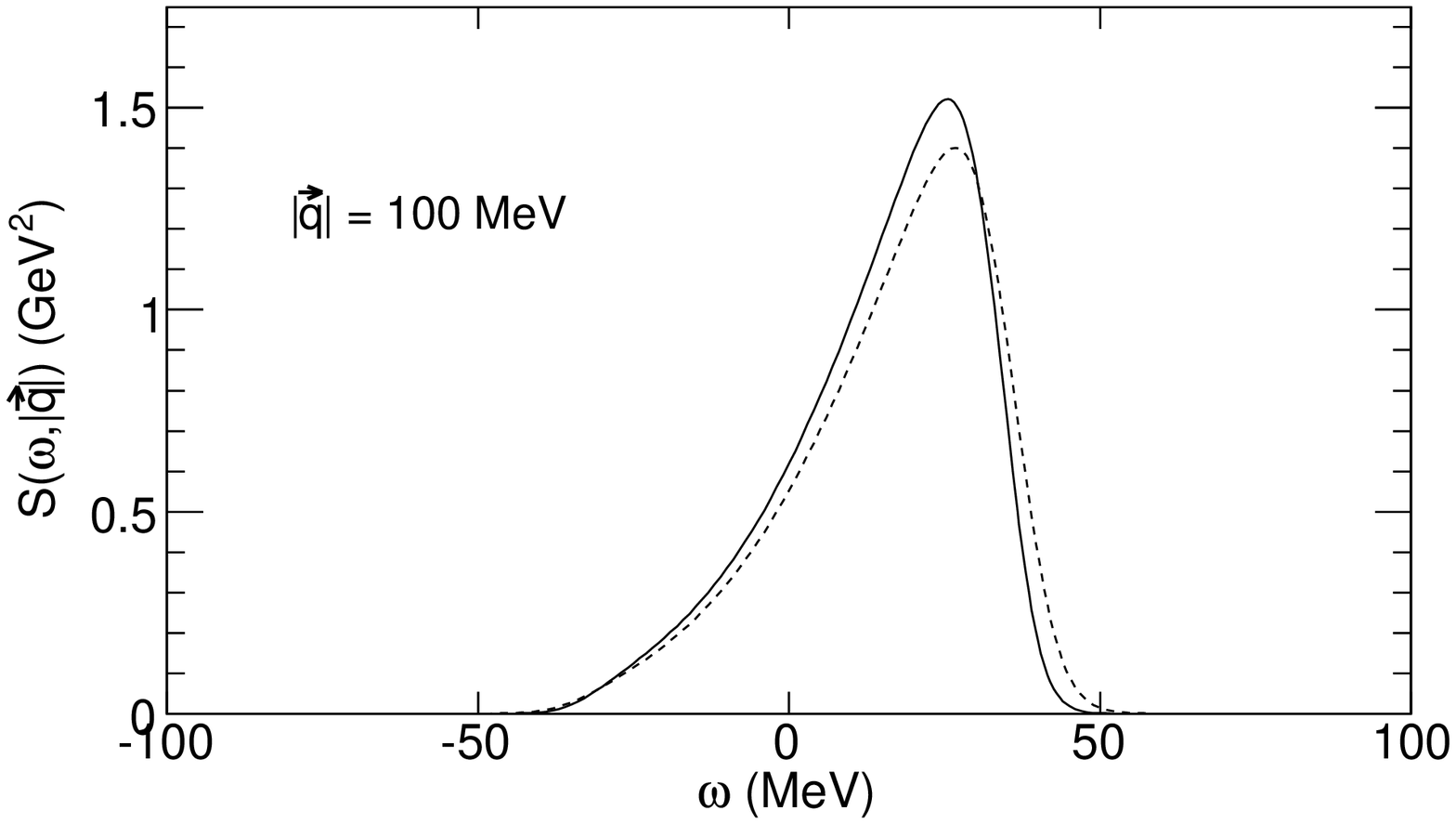}\\
\includegraphics[width=0.49\textwidth]{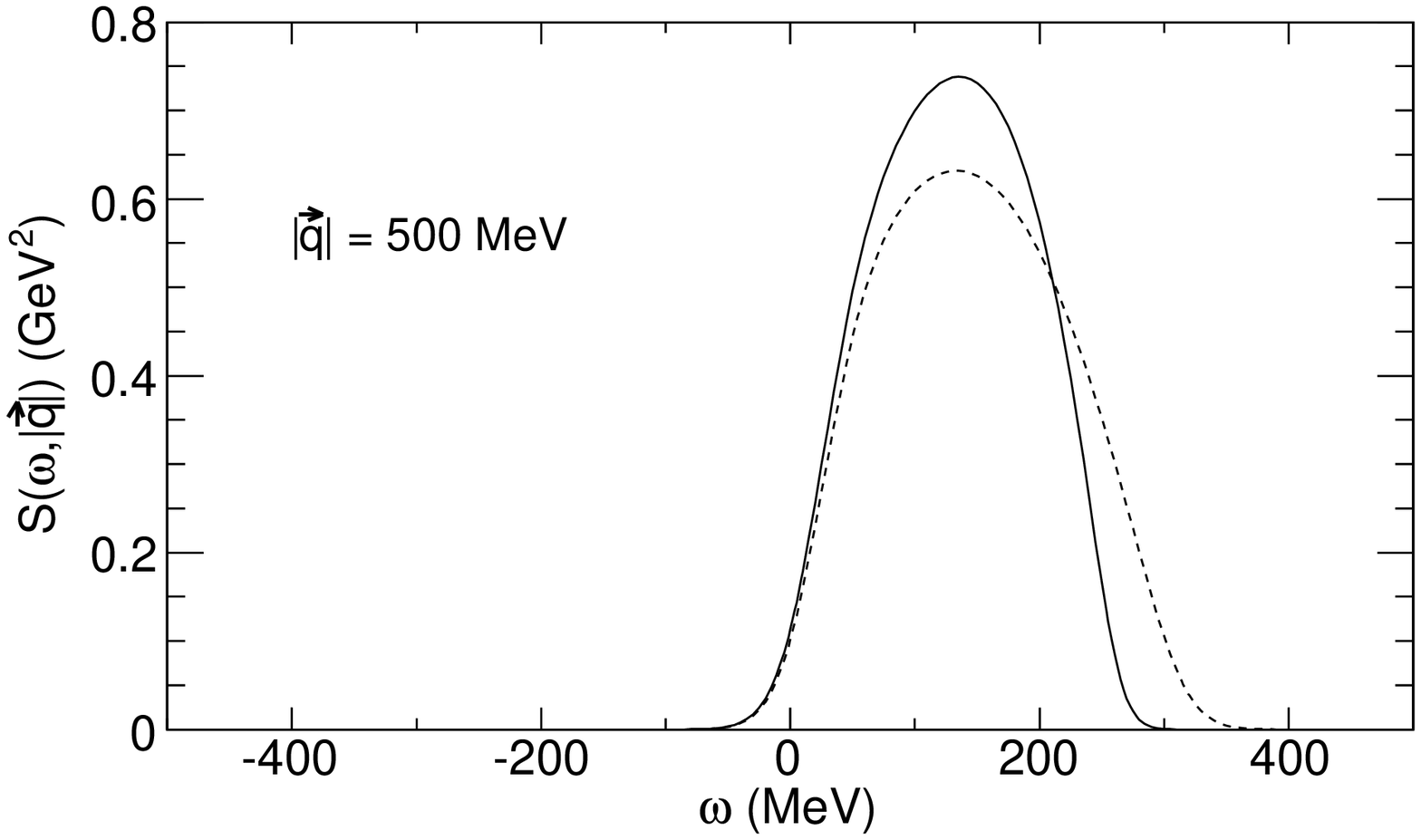}\hspace{\stretch{1}}
\includegraphics[width=0.49\textwidth]{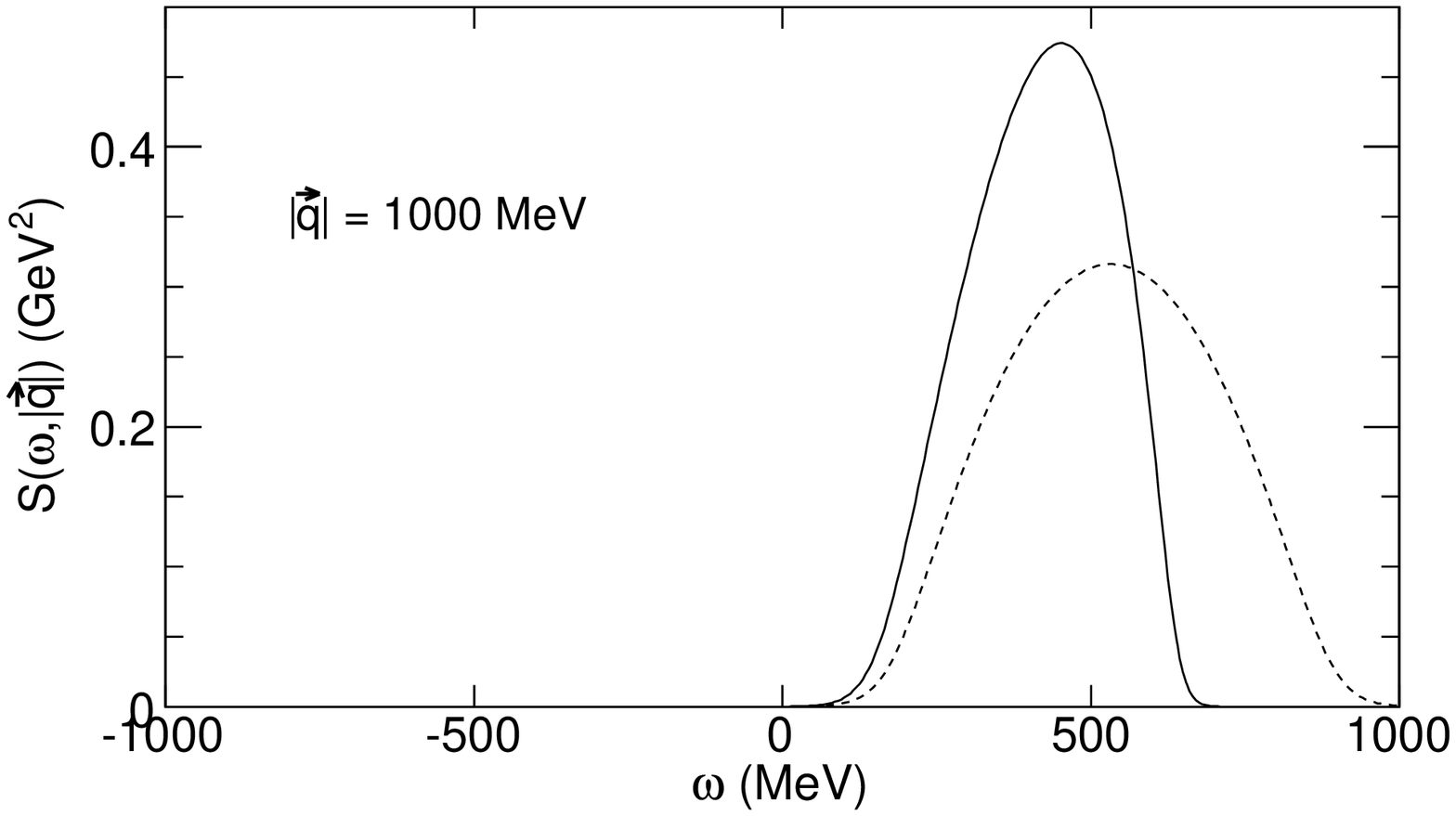}
\caption{The dynamical form factor $S(\omega,|\vec{q}|)$ for a neutron
gas with density $n = 0.16$ fm$^{-3}$ at a temperature of $kT = 10$
MeV, for various values of the momentum transfer. The solid line
represents the relativistic calculation, the dashed line the
non-relativistic one.}\label{dynamical}
\end{figure*}

First, we investigate the role of relativistic effects in the
phase-space factor. To this end, it is illustrative to introduce 
\begin{widetext}
\begin{equation}\label{eq411}
\frac{d^3\Phi}{d|\vec{p}_x|d^2\Omega_x} =
\sum_{\substack{\alpha_a,\alpha_b,\\ \alpha_x,\alpha_y}}\int
F^b_{\alpha_b}(\vec{p}_b)d^3\vec{p}_b\:\int
F'^y_{\alpha_y}(\vec{p}_y)\:
d^3\vec{p}_y  \delta^4(p_a+p_b-p_x-p_y) 
F'^x_{\alpha_x}(\vec{p}_x)|\vec{p}_x|^2 \; .
\end{equation}
\end{widetext}
This quantity is independent of the dynamics of the $a + b \rightarrow
x + y $ reaction.  Even at low neutrino energies, sizable differences between
relativistic and non-relativistic expressions for the phase-space
integral emerge.  In order to quantify this, 
we compare the dynamic form factor $S(\omega,|\vec{q}|)$ of the 
non-relativistic calculation of Ref.~\cite{Reddy:1997yr} with our relativistic
one.  Defining $w = q = p_a - p_x$ and assuming that $m_y = m_b$,
Eq.~\eqref{eq411} can be cast in the following form
\begin{equation}
\frac{d^3\Phi}{d|\vec{p}_x|d^2\Omega_x} = F'^{x}_{\alpha_x}(\vec{p}_x)
|\vec{p}_x|^2S(\omega,|\vec{q}|).
\end{equation}
As shown in Ref.~\cite{Reddy:1997yr}, in the non-relativistic limit
this equation has an analytical solution.  For completeness we wish to
mention that even in the relativistic situation a solution for
$S(\omega,|\vec{q}|)$ in terms of polylogarithmic functions can be found
in Ref.~\cite{Grypeos:1998zt}.  From Fig.~\ref{fermicompare}, it is
clear that above a certain momenta, the relativistic Fermi
distribution $F_R$ is larger than its non-relativistic counterpart $F_{NR}$. In
general,
\begin{equation}
\int\limits_{p_-}^\infty F_{R}(p) p^2dp \ge \int\limits_{p_-}^\infty F_{NR}(p) p^2dp.
\end{equation}
As can be appreciated from Fig.~\ref{dynamical}, this relation ensures
that $S(\omega,|\vec{q}|)$ is larger when computed in a relativistic
approach.  At moderate momentum transfers, the effect of relativity is
limited to affecting the absolute magnitude of $S(\omega,|\vec{q}|)$.
At higher $|\vec{q}|$-values, an additional effect in the $\omega$
dependence of $S(\omega,|\vec{q}|)$ appears.  The shift towards lower
values of $\omega$ is related to
the differences in the lower limit of the integration over
$|\vec{p}_b|$ in $S(\omega,|\vec{q}|)$.  For higher momentum exchange,
the term $F'^y_{\alpha_y}(\vec{q}+\vec{p}_b)$ in Eq.~\eqref{eq411}
approaches one. 
Hence the influence of the momentum distribution of the
gas $F_{\alpha_b}^b(\vec{p}_b)$ becomes prominent, causing $S(\omega,|\vec{q}|)$ to reach a maximum at $p_b^{MIN} = 0$.
Relativistically, the corresponding value for the energy exchange $\omega$ is given by $q^2/(m_b+\sqrt{m_b^2+q^2})$,
while in a non-relativistic calculation this quantity equals
$q^2/(2m_b)$.

The role of relativity in the dynamical form factor
$S(\omega,|\vec{q}|)$ can be easily estimated in the non-degenerate
limit. Indeed, the classical limit will be reached when $(\mu_i-M)/kT
\ll 0$. Under those conditions, one obtains the following closed
relativistic form for $S(\omega,|\vec{q}|)$
\begin{widetext}
\begin{equation}
S(\omega,|\vec{q}|) = \frac{2\pi}{|\vec{q}|}
kT\left[\omega(m_b+kT+T_-) + (m_b+kT+T_-)^2 + (kT)^2 \right]
\exp\left(\frac{\mu_b-E_-}{kT} \right)  \; ,
\end{equation}
\end{widetext}
where the following quantities were introduced,
\begin{equation}
E_- = \sqrt{{p_b^{MIN}}^2 + m_b^2}, \qquad T_- = E_- - m_b,
\end{equation}
A non-relativistic approach leads to a closed expression with only one
term \cite{Reddy:1997yr}
\begin{equation}
S(\omega,|\vec{q}|) = \frac{2\pi}{|\vec{q}|}kT
m_b^2\exp\left(\frac{\mu_b-e_-}{kT}\right) \; ,\label{SNR}
\end{equation}
with,
\begin{equation}
e_- = \frac{1}{4}\frac{(\omega-|\vec{q}|^2/2m_b)^2}{|\vec{q}|^2/2m_b}.
\end{equation}
The main difference between the relativistic and non-relativistic
  expressions for $S(\omega,|\vec{q}|)$ in the classical limit, is the dependence on
  $kT$ and $T_-$, and the presence of a term proportional to $\omega$
  in the relativistic formulation.  It can be shown that
  Eq.~\eqref{SNR} is valid when $|\vec{q}|\ll 2m_b$ and $\omega\approx
  0$.  For $|\vec{q}|\ll 2m_b$ and finite values of $\omega$, a correction given by the
  $(m_b+kT+T_-)^2$ is essential. When $|\vec{q}|\ll 2m_b$ is not valid
  anymore, the second correction $\omega(m_b+kT+T_-)$ starts playing
  an important role.  The term $(kT)^2$ only becomes important at higher temperatures.

\subsection{Cross sections}\label{sec4:2}

\begin{figure}
\includegraphics[scale=0.45]{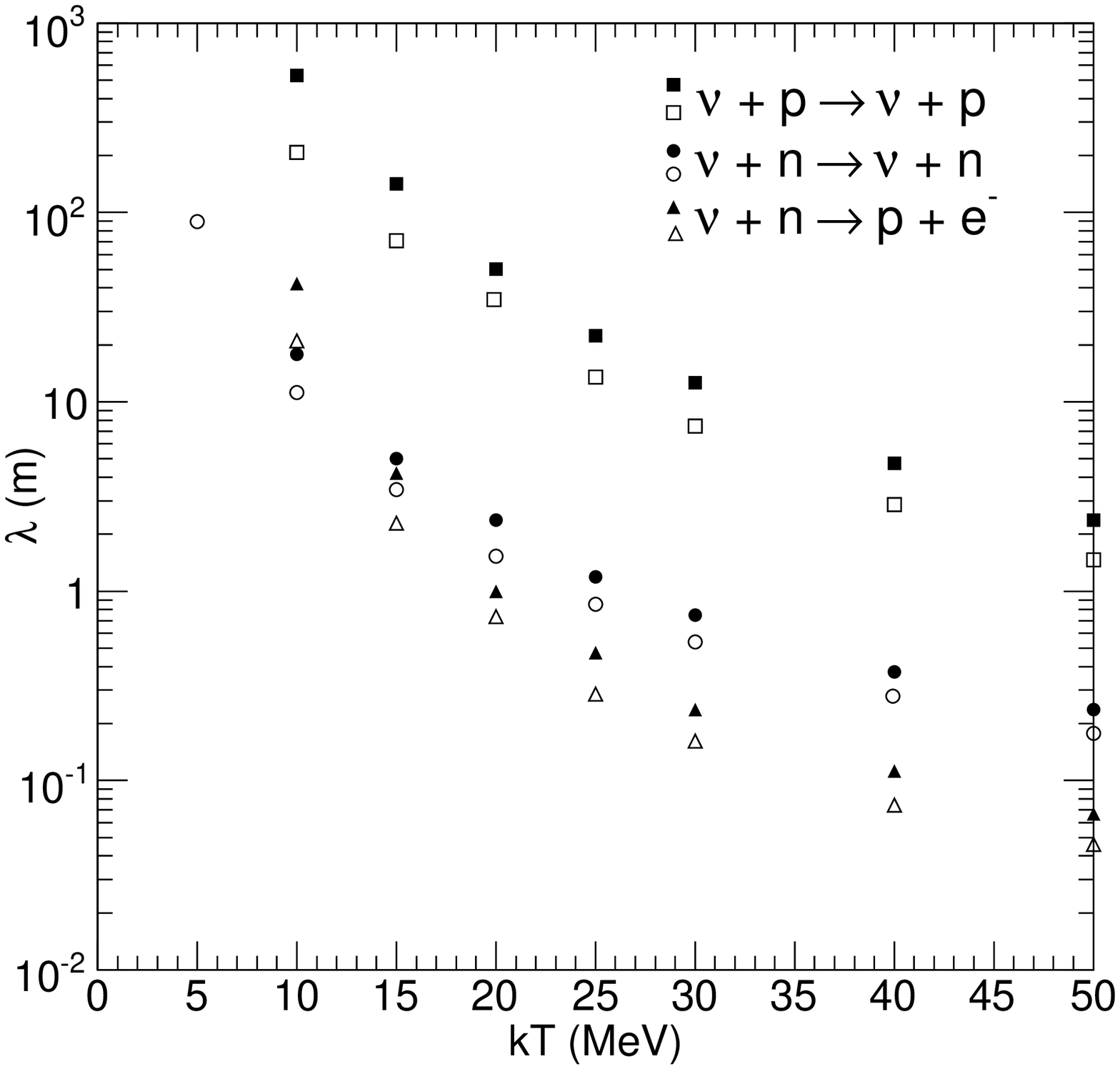}
\caption{Neutrino mean free path as a function of the temperature for
a nucleon gas in beta equilibrium with $\mu_\nu = 0$ and $n = 0.16$
fm$^{-3}$.  All results are for an initial neutrino energy of 
$3kT$. The open markers present results of our relativistic
calculation and are compared to the non-relativistic results of
Ref.~\cite{Reddy:1997yr} (filled markers). The squares represent the mean free path for
neutral-current scattering off protons ($\nu +p \rightarrow \nu + p$),
circles for neutral-current interactions with neutrons ($\nu+ n\rightarrow\nu+n$),
and the triangles for the charged-current neutrino-interaction
($\nu+n\rightarrow p+e^-$). }
\label{rel_nrel}
\end{figure}

The differences between relativistic and non-relativistic calculations
of the dynamical form factor translate themselves in the corresponding
cross sections.  Therefore, we studied the influence of a relativistic
treatment of the Fermi function on the cross sections and the neutrino
mean free path $\lambda=\frac{1}{n\sigma}$ .  Fig.~\ref{rel_nrel}
shows the difference in mean free path between our relativistic and
the non-relativistic calculation of Ref.~\cite{Reddy:1997yr}, for
neutral and charged-current processes.  It is clear that relativistic
calculations predict larger cross sections and a smaller mean free
path for the neutrinos.

\begin{figure}
\includegraphics[scale=0.45]{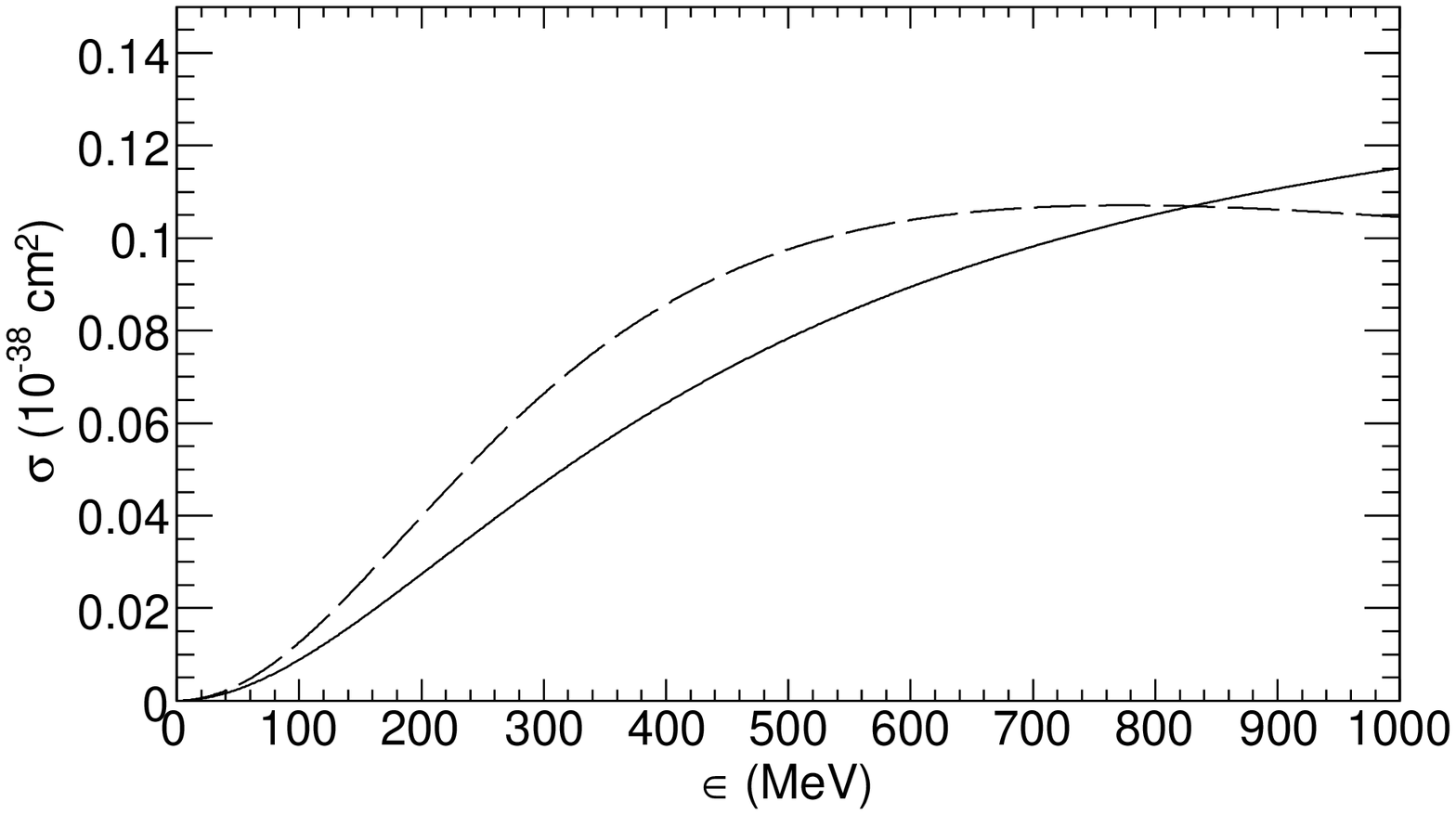}
\caption{Polarization of the outgoing nucleon for neutral-current
scattering off a neutron.  The spin-quantization axis is defined
parallel with the direction of the incoming neutrino.  The full (dashed) line
represents the contribution for final nucleons with spin-up
(spin-down) to the cross sections.}
\label{onenucl_spin}
\end{figure}
  
\begin{figure*}
\includegraphics[width=0.49\textwidth]{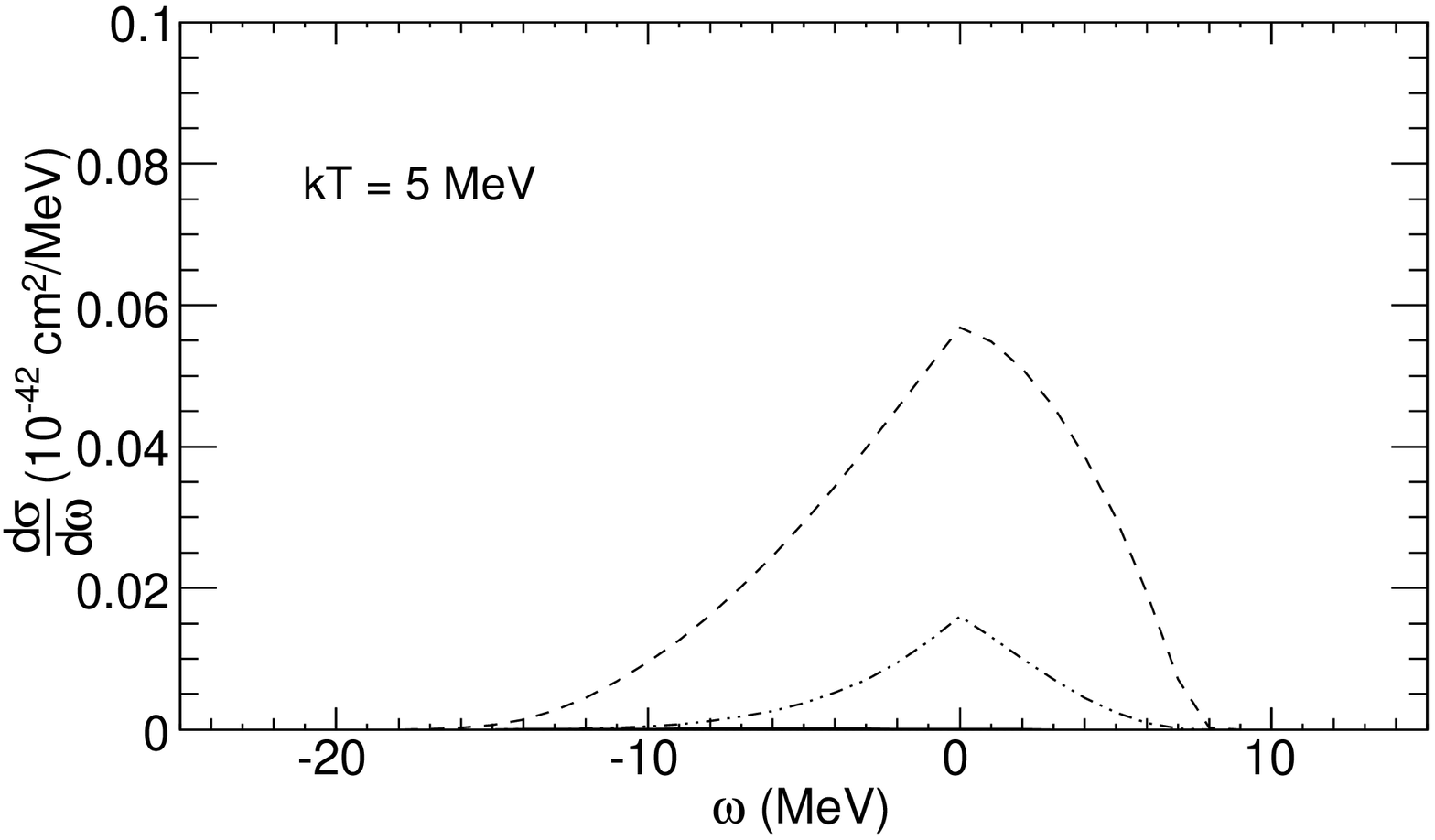}\hspace{\stretch{1}}
\includegraphics[width=0.49\textwidth]{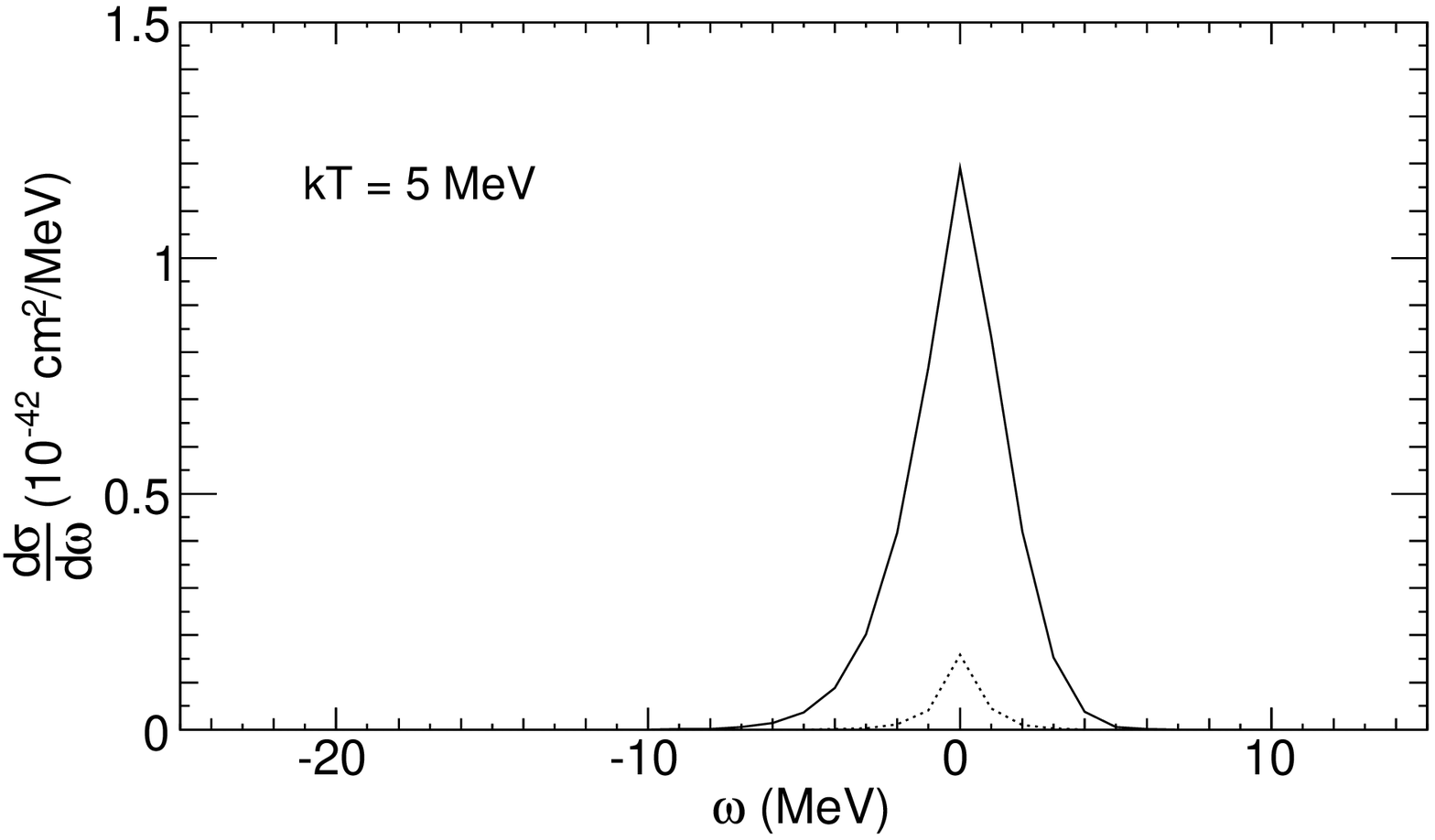}\\
\includegraphics[width=0.49\textwidth]{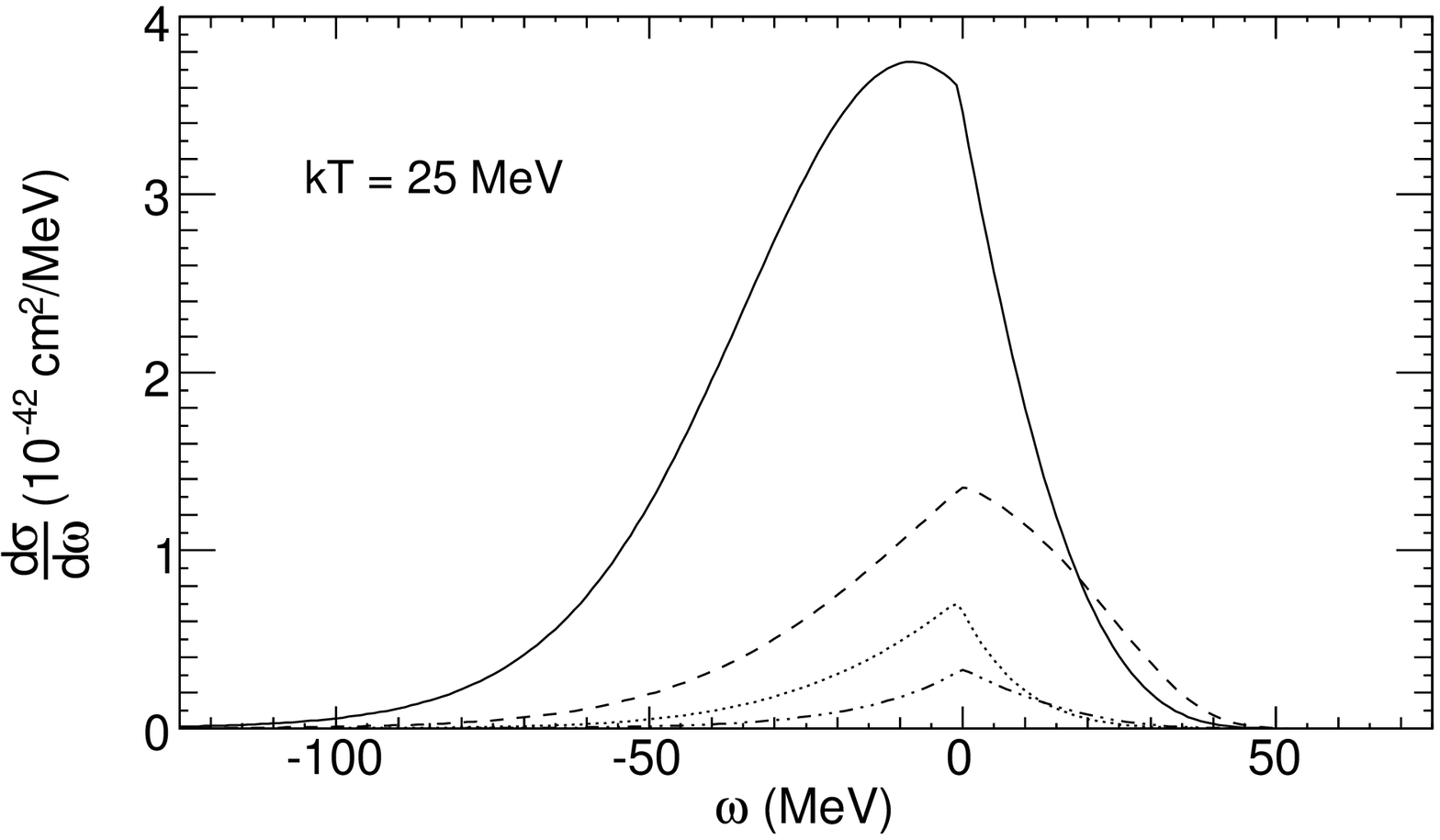}\hspace{\stretch{1}}
\includegraphics[width=0.49\textwidth]{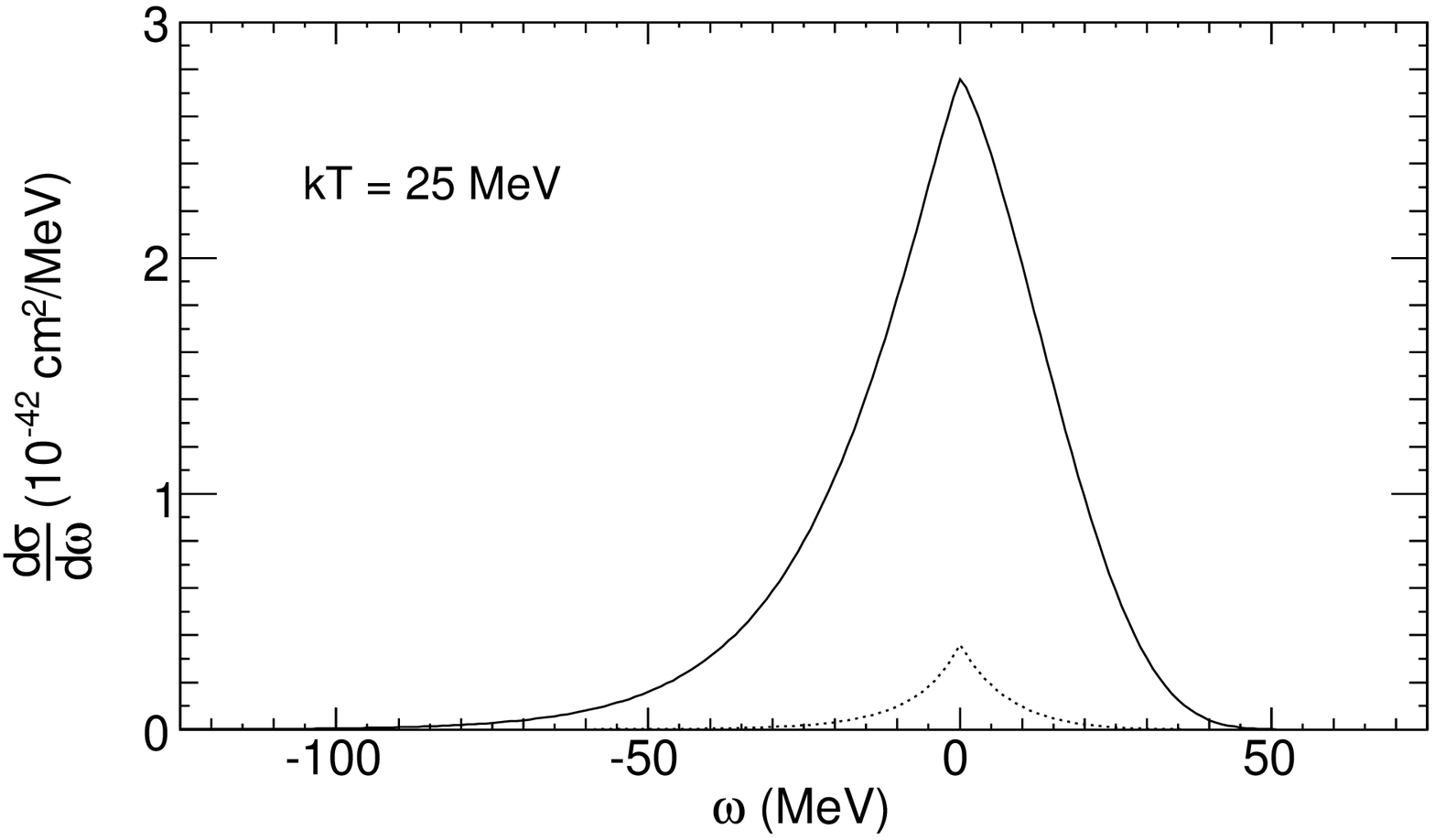}\\
\includegraphics[width=0.49\textwidth]{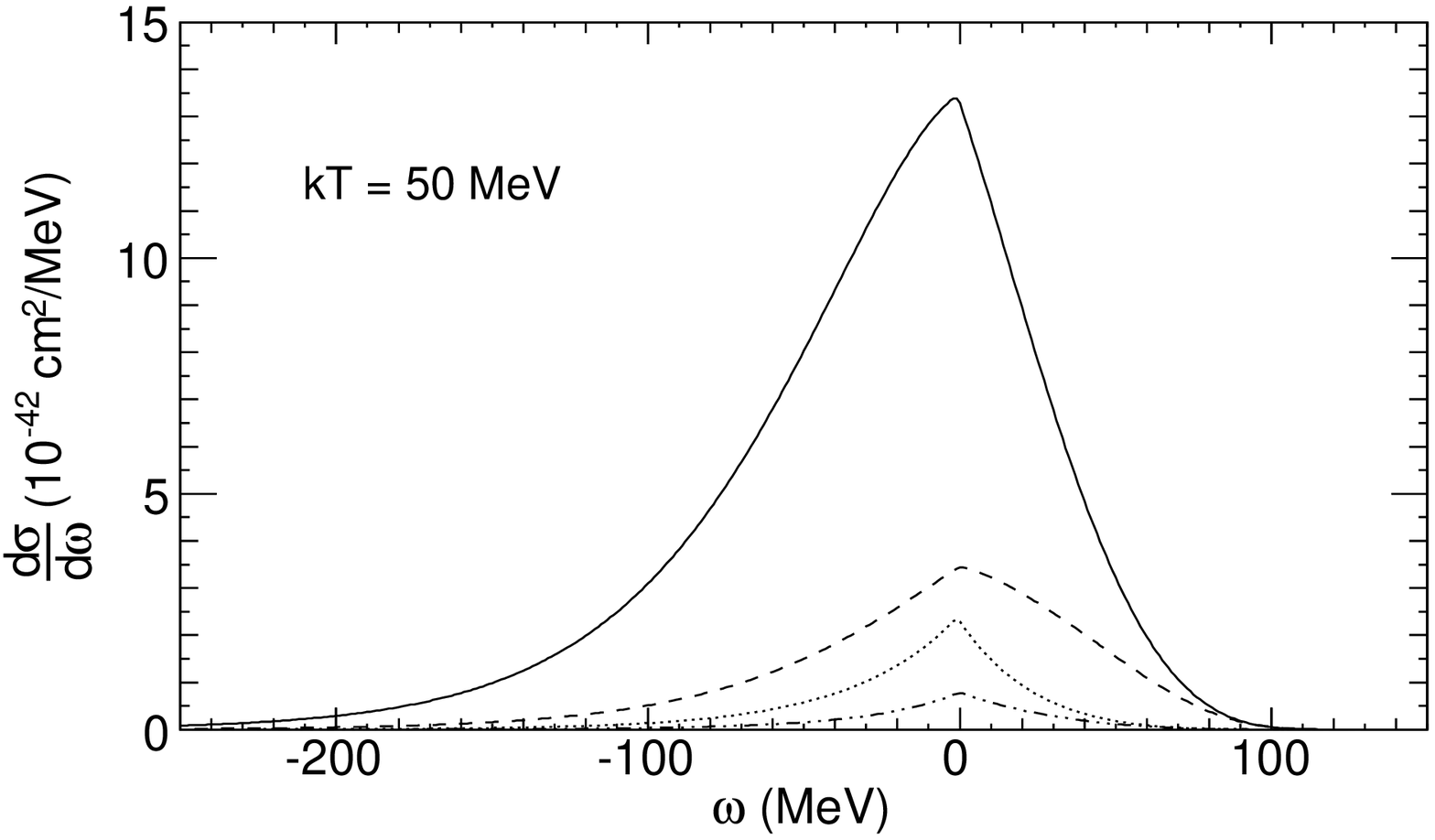}\hspace{\stretch{1}}
\includegraphics[width=0.49\textwidth]{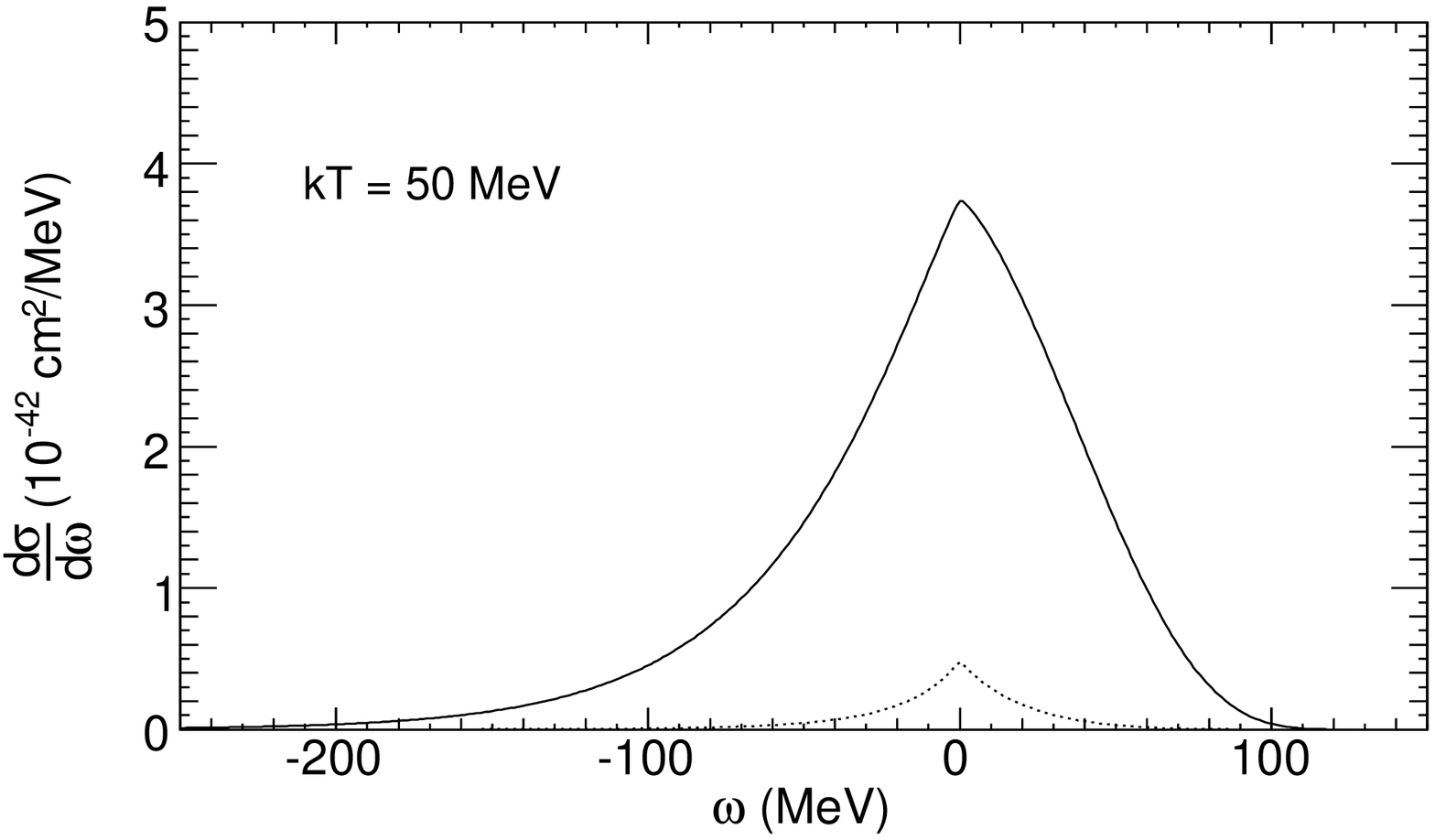}
\caption{Different polarization contributions to the differential
cross section for neutrino scattering in neutron matter with a density
$n_B = 0.16$ fm$^{-3}$ and temperatures $kT = 5,\;25,\;50$ MeV.  The
incoming neutrino energy is $3kT$.  The left panels present neutron
cross sections: CC with final spin down (solid line) ; CC with final spin up (dotted) ; NC
with final spin down (dashed) ; NC with final spin up (dashed-dotted). The right
panels are for neutrino-proton scattering : NC with final spin down
(solid line) and NC with final spin up (dotted).  The quantization
axis for the spin of the outgoing nucleon was chosen parallel to the
incoming neutrino's momentum.}\label{X}
\end{figure*}

Figs.~\ref{onenucl_spin} and \ref{X} illustrate another interesting
feature of the neutrino-nucleon processes. The weak magnetic
contribution in the current operator, which was earlier shown to play
a substantial role, induces some asymmetry for the polarization of the
final nucleon.  Figures \ref{onenucl_spin} and \ref{X} display the
separated spin contributions to the cross sections as a
function of the incoming neutrino energy and energy exchange.  Cross
sections are shown for both polarization directions of the final
nucleon. Thereby, the spin quantization axis is along the direction of the impinging neutrino.
The cross section for a final nucleon with $S = - \frac{1}{2}$ is 
considerably larger than the one for $S = + \frac{1}{2}$.

Polarization effects were shown to be large in neutrino-induced
nucleon knockout reactions from nuclei \cite{Jachowicz:2004we,Jachowicz:2005np,Lava:2005pb}. 
Apparently,  similar mechanisms govern the spin dependence of neutrino
interactions in nuclear matter.  The polarization effects might be of
importance under circumstances of strong magnetic
fields.  In Ref.~\cite{Duan:2005fc}, a study of the neutrino-nucleon interactions in a
strong magnetic field is presented. Thereby, the weak magnetic term
responsible for sizable spin-polarization asymmetries, has been neglected.

The strong suppression of charged-current reactions at low incoming
neutrino energies, is mainly due to the Pauli blocking of the outgoing 
electrons.  This is seen in the top left panel of Fig.~\ref{X} for $kT = 5$ MeV.  The CC cross section is smaller by several orders of magnitude than the NC one. For a hadron gas at low
temperatures in beta equilibrium, the chemical potential is about 60
MeV for a density of 0.16 fm$^{-3}$.  As a consequence, low energetic
neutrinos are not able to create an electron which is not blocked.
Fig.~\ref{X} shows clearly that the charged-current channel is the
most important one at higher energies. For a neutrino energy of 150
MeV, the chemical potential is 114 MeV. Higher temperatures make the
Fermi distribution shift to higher energies, an effect further
enhancing the charged-current cross sections.

Summarizing, we studied neutrino interactions in a relativistic Fermi
gas.  Our results for neutrino scattering on a single nucleon are in
excellent agreement with previous calculations.  In the limit of low
temperatures and vanishing nucleon chemical potential, our cross-section
calculations coincide with those for scattering on a single nucleon.
We find important differences in neutrino cross sections and the
neutrino mean free path between relativistic and non-relativistic
calculations.  In a relativistic treatment, neutrino cross sections
are larger.  We have shown that the weak magnetic contribution in the
weak current is responsible for considerable asymmetries in the
polarization of the scattered nucleons. In future work, the
correlations will be implemented.

\begin{acknowledgments}
The authors wish to thank H.-T.~Janka for interesting discussions.
They are grateful to the Fund for Scientific Research (FWO) Flanders
and the UGent University Research Board for financial support.
\end{acknowledgments}

\end{document}